\documentclass[5p,number,sort&compress]{elsarticle}

\usepackage{graphicx}
\usepackage{color}%
\usepackage{amsmath}
\usepackage{amssymb}
\usepackage{mathtools}
\usepackage{bm}
\usepackage{accents}
\usepackage{dcolumn}
\usepackage{multirow}
\usepackage{arydshln}
\usepackage{placeins}
\usepackage{hyperref}
\definecolor{link}{rgb}{0.07, 0.07, 0.80}
\hypersetup{
  pdftitle={Phase-space wavepacket dynamics of internal conversion via conical intersection: Multi-state quantum Fokker-Planck equation approach},
  pdfauthor={Tatsushi IKEDA, Yoshitaka TANIMURA},
  colorlinks=true,
  linkcolor=link,
  citecolor=link,
  urlcolor=link
}

\makeatletter
\newcommand*{\nolink}[1]{{\begin{NoHyper}#1\end{NoHyper}}}
\newcommand{\manuallabel}[2]{\def\@currentlabel{#2}\label{#1}}
\makeatother

\begin{document}
\manuallabel{sec:s-qfpe-derivation}{S1}
\manuallabel{sec:s-adiabatic-rep}{S2}
\manuallabel{sec:s-brinkman-derivation}{S3}
\manuallabel{sec:s-adiabatic}{S4}
\manuallabel{sec:s-c-misc}{S5}
\manuallabel{sec:s-artificial-viscosity}{S6}
\manuallabel{sec:s-negative-F_e}{S7}

\title{Phase-space wavepacket dynamics of internal conversion via conical intersection: Multi-state quantum Fokker-Planck equation approach}

\author[kyoto-u]{Tatsushi Ikeda}
\ead{ikeda.tatsushi.37u@kyoto-u.jp}
\author[kyoto-u]{Yoshitaka Tanimura}
\ead{tanimura.yoshitaka.5w@kyoto-u.jp}

\address[kyoto-u]{Department of Chemistry, Graduate School of Science, Kyoto University, Kyoto 606-8502, Japan}

\date{\today}

\begin{abstract}
  We theoretically investigate internal conversion processes of a photoexcited molecule in a condensed phase.
  The molecular system is described by two-dimensional adiabatic ground and excited potential energy surfaces that are coupled to heat baths.
  We quantify the role of conical intersection (CI) and avoided crossing (AC) in the PESs in dissipative environments by simulating the time evolution of wavepackets to compute the lifetime of the excited wavepacket, yield of the product, and adiabatic electronic coherence.
  For this purpose, we employ the multi-state quantum Fokker-Planck equation (MSQFPE) for a two-dimensional Wigner space utilizing the Wigner-Moyal expansion for the potential term and the Brinkman hierarchy expression for the momentum.
  We find that the calculated results are significantly different between the CI and AC cases due to the transition in the tuning mode and vibrational motion in the coupling mode.
\end{abstract}


\maketitle

\section{Introduction}
Internal conversion (IC) of a photoexcited molecule through conical intersection (CI) points plays an important role in photochemical processes, such as ultrafast electronic and vibrational transitions \cite{domcke2004book, domcke2011book, koppel1984acp, domcke1997acp, gonzalez2000pnas, worth2004arpc, polli2010nature, hamm2012prl, tahara2016jpca, prokhorenko2016jpcl, dijkstra2017jcp, kowalewski2017cr, thoss2000jcp, kuhl2002jcp, chen2016fd, miyata2017mc}.
Such systems are characterized by adiabatic ground and excited potential energy surfaces (PESs) that are functions of multi-dimensional reaction coordinates.
Conical intersections arise at the intersections of the PESs, where the Born-Oppenheimer (BO) approximation breaks down.
As a result, efficient non-adiabatic transitions occur through CIs.
Experimentally, the roles of CIs are not easy to explore, because excited wavepackets at the CI points are in transition states exhibiting very short lifetimes.
Fort this reason, theoretical input regarding de-excitation processes via CIs are important for analyzing such experimental results.

Theoretically, investigations of this kind are carried out by obtaining the profiles of the multi-dimensional PESs and then performing quantum kinetic simulations running the wavepackets in the PESs.
However, the existence of singular points at CI points make it difficult to obtain accurate static structures of the PESs and to carry out reliable kinetic simulations.
For kinetic simulations, because the momentum and lifetime of the wavepackets in excited states are also determined by thermal activation and relaxation processes,  environmental effects must be described in terms of open quantum dynamics; quantum mechanically consistent treatments of the system and environment are essential to obtain reliable results.
The excited state dynamics of such systems that explicitly take into account nuclear degrees of freedom and electronic states have been investigated using approaches that employ equations of motion for the density matrices or phase space distributions \cite{thoss2000jcp, kuhl2002jcp, chen2016fd, miyata2017mc, cotton2017jcp, tanimura1994jcp, tanimura1997jcp, maruyama1998cpl, ikeda2017jcp, kapral1999jcp, ando2003jcp, hanna2005jcp}, mixed quantum-classical trajectories \cite{delos1972pra, tully1990jcp, hammes1994jcp, coker1995jcp, santer2001jcp, subotnik2016arpc}, and Gaussian quantum wavepackets \cite{ben1998jcp, ben2000jcpa, makhov2014jcp}.
Many of these approaches, however, involve non-trivial assumptions, in particular, those regarding the quantum dynamical treatment of the couplings between electronic states and reaction coordinates, without careful verification of their validity.

\begin{figure}
  \centering
  \includegraphics[scale=0.9]{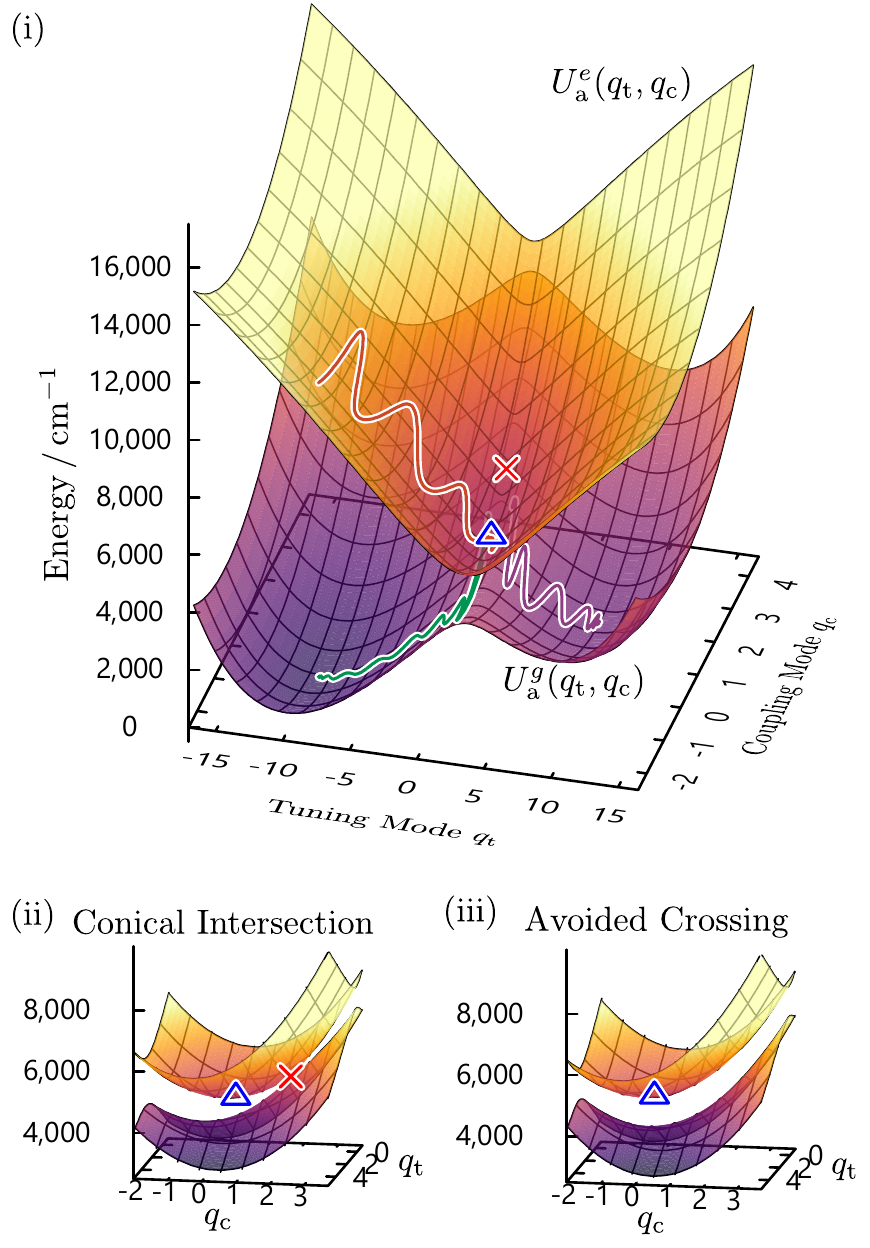}
  \caption{ Adiabatic BO PESs for our models.
    (i) The adiabatic ground and excited BO PESs, $U_{\mathrm{a}}^{g}(\vec{q})$ and $U_{\mathrm{a}}^{e}(\vec{q})$, are depicted for the CI1 model with $d=1.5$.
    The red, green, and purple curves represent the mean coordinates $\langle \vec{q}(t)\rangle ^{e}$ for the excited state, $\langle \vec{q}(t)\rangle ^{0}$ for the reactant region of the ground state, and $\langle \vec{q}(t)\rangle ^{1}$ for the product region of the ground state, respectively, for the photo-excited process.
    The symbols $\times $ and $\triangle $ represent the CI point and the minimum of $U_{\mathrm{a}}^{e}(\vec{q})$, respectively.
    (ii) The right-side view of (i).
    (iii) The right-side view of the crossing region of the AC1 model with $a=500~\mathrm{cm}^{-1}$.
  }
  \label{fig:PES}
\end{figure}

In this paper, we investigate the role of CI and avoided crossing (AC) under dissipative environments.
Although the qualitative analysis to investigate the role of them has been carried out \cite{blazer2003cpl}, we show a quantitative analysis by controlling diabatic coupling parameters.
For this purpose, we employ the multi-state quantum Fokker-Planck equation (MSQFPE) approach, which is an extension of the quantum Fokker-Planck equation for a Wigner distribution function (WDF) to a multi-state system \cite{tanimura1994jcp, tanimura1997jcp, maruyama1998cpl, ikeda2017jcp}.
While this technique allows us to treat systems with arbitrary PESs by taking into account the electric coherence explicitly, solving the MSQFPE is numerically demanding, in particular, in the case of multi-dimensional PESs.
For this reason, we apply the Wigner-Moyal expansion to a multi-state Wigner distribution function (MSWDF) in order to evaluate the quantum Liouvillian for the multi-state system and the Brinkman hierarchy for the momentum space of the MSWDF.
Then, we explore the interplay between thermal relaxation and de-excitation dynamics via the CI and AC by studying wavepackets dynamics on PESs (see Fig.~\ref{fig:PES}).

We employed a general-purpose graphics processing unit (GPGPU) to carry out the numerical integrations of the MSQFPE in this work.

The organization of this paper is as follows:
In Sec.~\ref{sec:method}, we explain the features of a Hamiltonian used to model a molecular system described by the PESs of the electronic states and introduce the multi-state quantum Fokker-Planck equation.
In Sec.~\ref{sec:models}, we present models for two-dimensional IC processes.
In Sec.~\ref{sec:results}, we present the results for wavepacket dynamics that we obtained using the method introduced in Sec.~\ref{sec:method} and the models presented in Sec.~\ref{sec:models}, and we analyze the computed wavepacket profiles.
Section~\ref{sec:conclusion} is devoted to concluding remarks.

\section{Theory}
\label{sec:method}

\subsection{Hamiltonian}
We consider a molecular system with multiple electronic states $\{|j\rangle \}$ coupled to $N$ dimensionless reaction coordinates $\vec{q}=(\dots ,q_{s},\dots )$, where $j$ represents the electronic diabatic states and $s$ is the index for the reaction coordinates.
Here and hereafter, we employ dimensionless coordinate and a dimensionless momentum defined in terms of the actual coordinates and momentum $\bar{q}_{s}$ and $\bar{p}_{s}$, as $q_{s} \equiv \bar{q}_{s}\sqrt {m_{s}\omega _{s}/\hbar }$ and $p_{s}\equiv \bar{p}_{s}/\sqrt {m_{s}\hbar \omega _{s}}$, where $\omega _{s}$ is the characteristic frequency and $m_{s}$ is the effective mass of the $s$th mode.
The system Hamiltonian is expressed as \cite{tanimura1994jcp, tanimura1997jcp, maruyama1998cpl, ikeda2017jcp}
\begin{align}
  \hat{H}(\vec{p},\vec{q})&=\sum _{s}\frac{\hbar \omega _{s}}{2}\hat{p}_{s}^{2}+\sum _{j,k}|j\rangle U^{jk}(\vec{q})\langle k|
  \label{eq:system-Hamiltonian},
\end{align}
where $\vec{p}=(\dots ,p_{s},\dots )$ represents the momentum conjugate to the coordinates.
Here, the nuclear and electronic operators are denoted by hats, and we omit direct products with the identity, $\otimes \hat{1}$, in the kinetic term.
The electronic potential is also expressed in  matrix form as $\{\bm{U}(\vec{q})\}_{jk}=U^{jk}(\vec{q})$, where the diagonal element $U^{jj}(\vec{q})$ and the off-diagonal element $U^{jk}(\vec{q})$ $(j\neq k)$ represent the diabatic PES of the $j$th electronic state and the diabatic coupling between the $j$th and $k$th states, respectively.
Note that the frequency $\omega _{s}$ is determined by the curvature of the potential $\bm{U}(\vec{q})$ as
\begin{align}
  \hbar \omega _{s}&\sim \left.\frac{\partial ^{2}}{\partial q_{s}^{2}}U^{j_{0}j_{0}}(\vec{q})\right|_{\vec{q}=\vec{q}\,_{0}^{j_{0}}},
\end{align}
where $j_{0}$ is a primary state of the vibrational dynamics and $\vec{q}\,_{0}^{j}$ is a local minimum of the primary PES.

\subsection{Quantum Liouvillian and Wigner distribution functions}
Under the canonical quantization of the dimensionless momentum and coordinates, given by
\begin{align}
  \hat{p}_{s}&\rightarrow \frac{1}{i}\frac{\partial }{\partial z_{s}}
  &\text{and}&
  &q_{s}&\rightarrow z_{s},
\end{align}
the state of the system at time $t$ is described using the density matrix $\bm{\rho }(\vec{z},\vec{z}\,',t)$, where $\rho ^{jj}(\vec{z},\vec{z}\,')$ and $\rho ^{jk}(\vec{z},\vec{z}\,')$ $(j\neq k)$ represent the population of $|j\rangle $ and the coherence between $|j\rangle $ and $|k\rangle $, respectively.
The time evolution of the system is described by the Liouville-von Neumann equation as
\begin{subequations}
  \begin{align}
    \begin{split}
      \frac{d}{dt}\bm{\rho }(\vec{z},\vec{z}\,',t)&=-\mathcal{L}(\vec{z},\vec{z}\,')\bm{\rho }(\vec{z},\vec{z}\,',t),
    \end{split}
  \end{align}
  where $\mathcal{L}(\vec{z},\vec{z}\,')$ is the quantum Liouvillian defined as
  \begin{align}
    \begin{split}
      \mathcal{L}(\vec{z},\vec{z}\,')\bm{\rho }(\vec{z},\vec{z}\,')&\equiv -i\sum _{s}\frac{\omega _{s}}{2}\left(\frac{\partial ^{2}}{\partial z_{s}^{2}}-\frac{\partial ^{2}}{\partial z_{s}'^{2}}\right)\bm{\rho }(\vec{z},\vec{z}\,')\\
      &\quad \quad +\frac{i}{\hbar }\Bigl[\bm{U}(\vec{z})\bm{\rho }(\vec{z},\vec{z}\,')-\bm{\rho }(\vec{z},\vec{z}\,')\bm{U}(\vec{z}\,')\Bigr].
    \end{split}
    \label{eq:liouvillian}
  \end{align}
  \label{eq:liouville-eq}
\end{subequations}

We now introduce the multi-state Wigner distribution function (MSWDF) for a multi-electronic state system \cite{tanimura1994jcp}:
\begin{align}
  \bm{W}(\vec{p},\vec{q},t)&\equiv \frac{1}{(2\pi )^{N}}\int \!d\vec{r}\,e^{-i\vec{p}\cdot \vec{r}}\bm{\rho }\left(\vec{q}{+}\frac{\vec{r}}{2},\vec{q}{-}\frac{\vec{r}}{2},t\right),
  \label{eq:wigner-transform}
\end{align}
where $\vec{q}\equiv (\vec{z}{+}\vec{z}\,')/2$ and $\vec{r}\equiv \vec{z}{-}\vec{z}\,'$.
Both $\vec{p}$ and $\vec{q}$ are now c-numbers in this phase space representation.
The distribution $\bm{W}(\vec{p},\vec{q},t)$ always satisfies the normalization condition
\begin{align}
  \sum _{j}\int \!d\vec{p}\,\int \!d\vec{q}\,W^{jj}(\vec{p},\vec{q},t)&=1.
  \label{eq:normalization}
\end{align}
The MSWDF is ideal for studying quantum transport systems, because it allows for the treatment of continuous systems, utilizing open boundary conditions and periodic boundary conditions.
In addition, the formalism can accommodate the inclusion of an arbitrary time-dependent external field \cite{tanimura1997jcp, maruyama1998cpl}.
Moreover, because we can compare quantum results with classical results obtained in the classical limit of the equation of motion for the WDF, this approach is effective for identifying purely quantum effects \cite{anatole2004job, kato2013jpcb, sakurai2014njp, cabrera2015pra}.

Using $\bm{W}(\vec{p},\vec{q},t)$, the equation of motion~\eqref{eq:liouville-eq} can be expressed in the Wigner-Moyal expansion form as
\begin{subequations}
  \begin{align}
    \frac{\partial }{\partial t}\bm{W}(\vec{p},\vec{q},t)&=-\mathcal{L}_{W}(\vec{p},\vec{q})\bm{W}(\vec{p},\vec{q},t),
  \end{align}
  where $\mathcal{L}(\vec{p},\vec{q})$ is the quantum Liouvillian for the MSWDF, defined as
  \begin{align}
    \begin{split}
      &\mathcal{L}_{W}(\vec{p},\vec{q})\bm{W}(\vec{p},\vec{q})\\
      &\equiv \sum _{s}\omega _{s}p_{s}\frac{\partial }{\partial q_{s}}\bm{W}\left(\vec{p},\vec{q}\right)\\
      &\quad \quad +\frac{i}{\hbar }\Bigl[\bm{U}\left(\vec{q}\right)\star \bm{W}\left(\vec{p},\vec{q}\right)-\bm{W}\left(\vec{p},\vec{q}\right)\star \bm{U}\left(\vec{q}\right)\Bigr],
    \end{split}
    \label{eq:qle-Liouvillian}
  \end{align}
  \label{eq:qle}
\end{subequations}
where the star operator, $\star $, represents the Moyal product defined as \cite{moyal1949mpcps, imre1967jmp}
\begin{align}
  \star &\equiv \exp \left[{\sum _{s}\frac{i}{2}\Bigl(\underaccent{\leftarrow }{\partial }_{q_{s}}\underaccent{\rightarrow }{\partial }_{p_{s}}-\underaccent{\rightarrow }{\partial }_{q_{s}}\underaccent{\leftarrow }{\partial }_{p_{s}}\Bigr)}\right].
  \label{eq:moyal-star}
\end{align}
Here we have introduced the differentiation operations from the left and right, which are defined by
\begin{align}
  \underaccent{\rightarrow }{\partial }_{x}f(x)=f(x)\underaccent{\leftarrow }{\partial }_{x}&\equiv \frac{\partial f(x)}{\partial x}.
\end{align}
The details of the derivation of Eq.~\eqref{eq:qle} are presented in Supplemental Information~\nolink{\ref{sec:s-qfpe-derivation}}.

Note that, as seen from Eqs.~\eqref{eq:qle-Liouvillian} and \eqref{eq:moyal-star}, the Moyal product itself does not explicitly involve the Planck constant $\hbar $ in its exponent when we adopt the dimensionless coordinates $\vec{q}$ and $\vec{p}$.
Higher-order terms can be omitted when the wavepackets in the momentum space are nearly Gaussian or the anharmonicity of the potential is weak.
In this case, Eq.~\eqref{eq:qle-Liouvillian} can be approximated by
\begin{align}
  \begin{split}
    &\frac{i}{\hbar }\Bigl[\bm{U}\left(\vec{q}\right)\star \bm{W}\left(\vec{p},\vec{q}\right)-\bm{W}\left(\vec{p},\vec{q}\right)\star \bm{U}\left(\vec{q}\right)\Bigr]\\
    &=\frac{i}{\hbar }\Bigl(\bm{U}(\vec{q})\bm{W}(\vec{p},\vec{q})-\bm{W}(\vec{p},\vec{q})\bm{U}(\vec{q})\Bigr)\\
    &\quad -\sum _{s}\frac{1}{2\hbar }\left(\frac{\partial \bm{U}(\vec{q})}{\partial q_{s}}\frac{\partial \bm{W}(\vec{p},\vec{q})}{\partial p_{s}}+\frac{\partial \bm{W}(\vec{p},\vec{q})}{\partial p_{s}}\frac{\partial \bm{U}(\vec{q})}{\partial q_{s}}\right).
  \end{split}
  \label{eq:pot-qcle}
\end{align}
The above is often used as the mixed quantum classical Liouville equation.
While verification of the validity of Eq.~\eqref{eq:pot-qcle} has been made in the large mass limit \cite{kapral1999jcp}, the validity of the numerical approximation in Eq.~\eqref{eq:moyal-star} must be examined case by case, because anharmonicity and the profiles of the wavepackets play roles in the approximation.
In the case of harmonic potentials, the above expression is exact \cite{tanimura1994jcp}, while we have to employ an integral convolution form of the potential terms in the full quantum case described by arbitrary PESs and diabatic coupling \cite{tanimura1997jcp, maruyama1998cpl, ikeda2017jcp}.

\subsection{The multi-state quantum Fokker-Plank equation}
We consider the situation in which each reaction coordinate $q_{s}$ interacts with its own heat bath.
This represents coupling to other inter- and intra-molecular modes.
The total Hamiltonian is then expressed as \cite{caldeira1983pa}
\begin{align}
  \begin{split}
    &\hat{H}_{\mathrm{tot}}(\vec{p},\vec{q},\vec{P},\vec{X})\\
    &=\hat{H}(\vec{p},\vec{q})+\sum _{s}\sum _{\xi }\frac{\hbar {\Omega _{s,\xi }}}{2}\left[\hat{P}_{s,\xi }^{2}+\left({X_{s,\xi }}-\frac{{g_{s,\xi }}}{{\Omega _{s,\xi }}}q_{s}\right)^{2}\right],
  \end{split}
\end{align}
where ${\Omega _{s,\xi }}$, ${X_{s,\xi }}$, and ${P_{s,\xi }}$ are the vibrational frequency, dimensionless coordinate, and its conjugate momentum of the $\xi $th bath mode coupled to $q_{s}$, respectively.
The coefficient ${g_{s,\xi }}$ is the coupling constant between $X_{s,\xi }$ and $q_{s}$.
Each bath is independent and is characterized by a spectral distribution function defined as
\begin{align}
  \mathcal{J}_{s}(\omega )&\equiv \sum _{\xi }\frac{g_{s,\xi }^{2}}{2}\delta (\omega -{\Omega _{s,\xi }}).
\end{align}
Here, we assume that each spectral distribution function $\mathcal{J}_{s}(\omega )$ has the Ohmic form
\begin{align}
  \mathcal{J}_{s}(\omega )&=\frac{\zeta _{s}}{\pi \omega _{s}}\omega ,
  \label{eq:ohmic}
\end{align}
where $\zeta _{s}$ is the system-bath coupling strength for the $s$th mode.
For the harmonic PES with frequency $\omega _{s}$, the conditions $\zeta _{s}<2\omega _{s}$, $\zeta _{s}=2\omega _{s}$, and $\zeta _{s}>2\omega _{s}$ for the $s$th mode correspond to the underdamped, critically damped, and overdamped cases, respectively.

The total system is described by $\bm{W}_{\mathrm{tot}}(\vec{p},\vec{q},\vec{P},\vec{X},t)$, where $\vec{X}=(\dots ,{X_{s,\xi }},\dots )$ and $\vec{P}=(\dots ,{P_{s,\xi }},\dots )$.
The reduced MSWDF is then given by
\begin{align}
  \bm{W}(\vec{p},\vec{q},t)&\equiv \int d\vec{P}\int d\vec{X}\,\bm{W}_{\mathrm{tot}}(\vec{p},\vec{q},\vec{P},\vec{X},t).
\end{align}
In general, the equations of motion for the reduced system are expressed in the hierarchical form \cite{tanimura1997jcp, maruyama1998cpl, ikeda2017jcp, tanimura1989jpsj, tanimura2006jpsj, tanimura2015jcp}.
In the Markovian case (i.e.~the case of an Ohmic distribution, Eq.~\eqref{eq:ohmic}, with the high-temperature approximation $\coth (\beta \hbar \omega _{s}/2)\sim 2/\beta \hbar \omega _{s}$, where $\beta $ is the inverse temperature divided by the Boltzmann constant, $\beta \equiv 1/k_{\mathrm{B}}T$), the equations of motion reduce to the quantum Fokker-Planck equation \cite{caldeira1983pa}.
In the present case, this is expressed as \cite{tanimura1994jcp}
\begin{align}
  \begin{split}
    \frac{\partial }{\partial t}\bm{W}(\vec{p},\vec{q},t)&=-\mathcal{L}_{W}(\vec{p},\vec{q})\bm{W}(\vec{p},\vec{q},t)\\
    &\quad +\sum _{s}\zeta _{s}\frac{\partial }{\partial p_{s}}\left(p_{s}+\frac{1}{\beta \hbar \omega _{s}}\frac{\partial }{\partial p_{s}}\right)\bm{W}(\vec{p},\vec{q},t).
  \end{split}
  \label{eq:qfpe}
\end{align}
The above equation is the MSQFPE, which is a generalization of the quantum Fokker-Planck equation for multiple modes and multiple electronic states.

\subsection{The adiabatic representation}
We now introduce electronic adiabatic states, $|\Phi ^{a}(\vec{q})\rangle $.
In the following, $a$, $b$, and $c$ refer to electronic adiabatic states and $j$, $k$, and $l$ refer to electronic diabatic states.
The $a$th electronic adiabatic state is an eigenfunction of the time-independent Schr\"odinger equation
\begin{align}
  \hat{U}(\vec{q})|\Phi ^{a}(\vec{q})\rangle &=U_{\mathrm{a}}^{a}(\vec{q})|\Phi ^{a}(\vec{q})\rangle 
  \label{eq:eigen},
\end{align}
where $\hat{U}(\vec{q})\equiv \sum _{j,k}|j\rangle U^{jk}(\vec{q})\langle k|$, and $U_{\mathrm{a}}^{a}(\vec{q})$ is the $a$th adiabatic BO PES.
The diabatic and adiabatic states are related through the transformation matrix expressed as
\begin{align}
  Z^{ja}(\vec{q})\equiv \langle j|\Phi ^{a}(\vec{q})\rangle 
  \label{eq:transformation-matrix}.
\end{align}
Using the matrix $\{\bm{Z}(\vec{q})\}_{ja}=Z^{ja}(\vec{q})$, Eq.~\eqref{eq:eigen} can be written in the form of a diagonal matrix as
\begin{align}
  \bm{Z}(\vec{q})^{\dagger }\bm{U}(\vec{q})\bm{Z}(\vec{q})&=\bm{U}_{\mathrm{a}}(\vec{q})
  \label{eq:diagonalization},
\end{align}
where $\{\bm{U}_{\mathrm{a}}(\vec{q})\}_{ab}\equiv \delta _{ab}U_{\mathrm{a}}^{a}(\vec{q})$, and $\delta _{ab}$ is the Kronecker delta.
The adiabatic representation of the (reduced) density matrix, $\bm{\rho }_{\mathrm{a}}(\vec{z},\vec{z}\,')$, is defined through application of the transformation matrix $\bm{Z}(\vec{q})$ to $\bm{\rho }(\vec{z},\vec{z}\,')$ as
\begin{align}
  \bm{\rho }_{\mathrm{a}}(\vec{z},\vec{z}\,')&\equiv \bm{Z}\left(\vec{z}\right)^{\dagger }\bm{\rho }(\vec{z},\vec{z}\,')\bm{Z}\left(\vec{z}\,'\right).
  \label{eq:a-density-matrix}
\end{align}
The adiabatic representation of the MSWDF, $\bm{W}_{\mathrm{a}}(\vec{p},\vec{q},t)$, is defined as
\begin{align}
  \bm{W}_{\mathrm{a}}(\vec{p},\vec{q},t)&\equiv \frac{1}{(2\pi )^{N}}\int \!d\vec{r}\,e^{-i\vec{p}\cdot \vec{r}}\bm{\rho }_{\mathrm{a}}\left(\vec{q}{+}\frac{\vec{r}}{2},\vec{q}{-}\frac{\vec{r}}{2},t\right),
  \label{eq:a-wigner-transform}
\end{align}
where the diagonal element $W_{\mathrm{a}}^{aa}(\vec{p},\vec{q},t)$ and the off-diagonal element $W_{\mathrm{a}}^{ab}(\vec{p},\vec{q},t)$ $(a\neq b)$ represent the population of $|\Phi ^{a}(\vec{q})\rangle $ and the coherence between $|\Phi ^{a}(\vec{q})\rangle $ and $|\Phi ^{b}(\vec{q})\rangle $, respectively.
\begin{subequations}
  The adiabatic representation of the MSWDF can also be obtained by applying $\bm{Z}(\vec{q})$ to $\bm{W}(\vec{p},\vec{q},t)$ as
  \begin{align}
    \bm{W}_{\mathrm{a}}(\vec{p},\vec{q},t)&=\bm{Z}\left(\vec{q}\right)^{\dagger }\star \bm{W}(\vec{p},\vec{q},t)\star \bm{Z}\left(\vec{q}\right).
    \label{eq:adiabatic-rep}
  \end{align}
  The inverse transformation is expressed as
  \begin{align}
    \bm{W}(\vec{p},\vec{q},t)
    &=\bm{Z}\left(\vec{q}\right)\star \bm{W}_{\mathrm{a}}(\vec{p},\vec{q},t)\star \bm{Z}\left(\vec{q}\right)^{\dagger }.
    \label{eq:adiabatic-rep-rev}
  \end{align}
  \label{eq:adiabatic-transform}
\end{subequations}
Details regarding Eqs.~\eqref{eq:transformation-matrix}, \eqref{eq:a-density-matrix}, \eqref{eq:adiabatic-rep}, and \eqref{eq:adiabatic-rep-rev} are presented in Supplemental Information~\nolink{\ref{sec:s-adiabatic-rep}}.

Integrating Eq.~\eqref{eq:qfpe} with the Liouvillian Eq.~\eqref{eq:qle}, we obtain the MSWDF in the diabatic representation, $\bm{W}(\vec{p},\vec{q},t)$.
Then, using the transformation~\eqref{eq:adiabatic-rep}, we obtain the MSWDF in the adiabatic representation, $\bm{W}_{\mathrm{a}}(\vec{p},\vec{q},t)$.
Although we can construct the MSQFPE for $\bm{W}_{\mathrm{a}}(\vec{p},\vec{q},t)$, numerical integration of this equation is difficult, because the present PESs in the adiabatic basis $|\Phi ^{a}(\vec{q})\rangle $ have a singularity at the CI point.
For this reason, in the present work, we solve Eq.~\eqref{eq:qfpe} for $\bm{W}(\vec{p},\vec{q},t)$ to obtain $\bm{W}_{\mathrm{a}}(\vec{p},\vec{q},t)$.

\subsection{Brinkman hierarchy}
While the quantum dynamics for an $N$-dimensional system can be studied by solving the Schr\"{o}dinger equation for $N$-dimensional wave functions, we must treat $2N$-dimensional reduced density matrices or Wigner functions in the study of open quantum dynamics evolving toward the thermal equilibrium state.
Although the present approach allows us to obtain accurate results, solving the MSQFPE is numerically demanding, in particular for multi-dimensional PESs.
Thus, to reduce the numerical cost, here we describe the momentum space of $\bm{W}(\vec{p},\vec{q},t)$ using a series expansion in terms of Hermite functions expressed as
\begin{subequations}
  \begin{align}
    \begin{split}
      \bm{W}(\vec{p},\vec{q},t)
      &=\sum _{\vec{n}\geq \vec{0}}e^{-\varepsilon \beta \bm{U}_{0}(\vec{q})/2}\bm{c}_{\vec{n}}(\vec{q},t)e^{-\varepsilon \beta \bm{U}_{0}(\vec{q})/2}\\
      &\quad \quad \times \psi _{\vec{0}}(\vec{p})\psi _{\vec{n}}(\vec{p}),
    \end{split}
    \label{eq:expansion}
  \end{align}
  where the function of the coordinates in matrix form, $\allowbreak \{\bm{c}_{\vec{n}}(\vec{q},t)\}_{jk}={c\,}_{\vec{n}}^{jk}(\vec{q},t)$, is defined as
  \begin{align}
    \begin{split}
      \bm{c}_{\vec{n}}(\vec{q},t)&\equiv \int \!d\vec{p}\,e^{+\varepsilon \beta \bm{U}_{0}(\vec{q})/2}\bm{W}(\vec{p},\vec{q},t)e^{+\varepsilon \beta \bm{U}_{0}(\vec{q})/2}\\
      &\quad \quad \times \psi _{\vec{n}}(\vec{p}){\psi _{\vec{0}}(\vec{p})}^{-1}.
    \end{split}
    \label{eq:expansion-rev}
  \end{align}
\end{subequations}
Here, the vector $\vec{n}\equiv (\dots ,n_{s},\dots )$ represents the momentum state of the Hermite function $\psi _{\vec{n}}(\vec{p})$, where $\psi _{\vec{n}}(\vec{p})\equiv \prod _{s}{\psi }^{(s)}_{n_{s}}(p_{s})$, with
\begin{align}
  \psi _{n_{s}}^{(s)}(p_{s})&\equiv \frac{1}{\sqrt {\mathstrut 2^{n_{s}}n_{s}!\alpha _{s}\sqrt {\mathstrut \pi }}}{H_{n_{s}}}\left(\frac{p_{s}}{\alpha _{s}}\right)\exp \left(-\frac{p_{s}^{2}}{2\alpha _{s}^{2}}\right).
\end{align}
Here, $\alpha _{s}\equiv \sqrt {\mathstrut 2/\beta \hbar \omega _{s}}$, and the $n$th Hermite polynomial is defined as ${H_{n}}(x)\equiv (-1)^{n}e^{x^{2}}(\partial ^{n}\!/\partial x^{n})e^{-x^{2}}$.
In Eqs.~\eqref{eq:expansion} and \eqref{eq:expansion-rev}, $\bm{U}_{0}(\vec{q})$ is the diagonal matrix whose diagonal elements are identical to those of $\bm{U}(\vec{q})$, and $\varepsilon $ is an arbitrary constant.
Because of Eq.~\eqref{eq:normalization}, $\bm{c}_{\vec{n}}(\vec{q},t)$ always satisfies
\begin{align}
  \sum _{j}\int \!d\vec{q}\,e^{-\varepsilon \beta U^{jj}(\vec{q})}{c\,}_{\vec{0}}^{jj}(\vec{q},t)=1.
\end{align}
The notation employed here is defined in Table~\ref{tab:multi-indice}.
\begin{table}
  \centering
  \caption{The multi-index notation we used in this paper.
    Note that $\vec{n}$ and $\vec{m}$ represent the multi-indexes $(\dots ,n_{s},\dots )$ and $(\dots ,m_{s},\dots )$, and $\vec{\alpha }$ and $\vec{q}$ are real vectors.
  }
  \begin{tabular}{cc@{~}c|cc@{~}c}
    \hline\hline
    Notation & \multicolumn{2}{c|}{Meaning}&
    Notation & \multicolumn{2}{c}{Meaning}\\
    \hline
    $ \vec{n}\leq \vec{m} $ &$\bigwedge _{s}$&$n_{s}\leq m_{s}$&
    $ \left|\vec{n}\right|$ & $\sum _{s}$&$n_{s}$\\
    $ \vec{\alpha }\,^{\vec{n}}$ & $\prod _{s}$&$\alpha _{s}^{n_{s}} $&
    $ \partial ^{\left|\vec{n}\right|}\!/\partial \vec{q}\,^{\vec{n}}$ & $\prod _{s}$&$\partial ^{n_{s}}\!/\partial q_{s}^{n_{s}}$\\
    $ \vec{n}\,!$ & $\prod _{s}$&$n_{s}!$&
    &\\
    \hline\hline
  \end{tabular}
  \label{tab:multi-indice}
\end{table}
Note that the zeroth order expansion term is proportional to the classical Boltzmann distribution: ${\psi }_{0_{s}}^{(s)}(p_{s}) \propto e^{-\beta \hbar \omega _{s}p_{s}^{2}/2}$.

The MSQFPE~\eqref{eq:qfpe} for $\bm{c}_{\vec{n}}(\vec{q},t)$ can be expressed in the form of simultaneous differential equations as
\begin{align}
  \begin{split}
    \frac{d}{dt}\bm{c}_{\vec{n}}\left(\vec{q},t\right)&=-\sum _{s}\mathcal{A}_{s}(\vec{q})\Bigl[\sqrt {\mathstrut n_{s}{+}1}\bm{c}_{\vec{n}{+}\vec{e}_{s}}(\vec{q},t)\\
      &\quad \quad \quad \quad \quad \quad \quad +\sqrt {\mathstrut n_{s}}\bm{c}_{\vec{n}{-}\vec{e}_{s}}(\vec{q},t)\Bigr]\\
    &\quad -\!\!\!\!{\sum _{\vec{0}\leq \vec{m}\leq \vec{n}}}\!\!\!\!\mathcal{B}_{\vec{m}}(\vec{q})\sqrt {\mathstrut \frac{\vec{n}\,!}{(\vec{n}{-}\vec{m})!}}\bm{c}_{\vec{n}{-}\vec{m}}(\vec{q},t)\\
    &\quad -\sum _{s}\zeta _{s}n_{s}\bm{c}_{\vec{n}}(\vec{q},t)
    \label{eq:brinkman-rep},
  \end{split}
\end{align}
where $\vec{e}_{s}$ is the $s$th unit vector and $\vec{m}\equiv (\dots ,m_{s},\dots )$ is the multi-index for the Moyal product~\eqref{eq:moyal-star}.
Here, we have introduced the superoperators $\mathcal{A}_{s}(\vec{q})$ and $\mathcal{B}_{\vec{m}}(\vec{q})$ defined as
\begin{subequations}
  \begin{align}
    &\begin{aligned}
      \mathcal{A}_{s}(\vec{q})\bm{c}_{\vec{n}}(\vec{q})&\equiv \frac{\alpha _{s}\omega _{s}}{\sqrt {\mathstrut 2}}\frac{\partial \bm{c}_{\vec{n}}(\vec{q})}{\partial q_{s}}\\
      &\quad -\frac{1}{\hbar }\Bigl[\bm{A}_{s}(\vec{q})\bm{c}_{\vec{n}}(\vec{q})+\bm{c}_{\vec{n}}(\vec{q})\bm{A}_{s}(\vec{q})\Bigr]\\
    \end{aligned}
    \intertext{and}
    &\begin{aligned}
      \mathcal{B}_{\vec{m}}(\vec{q})\bm{c}_{\vec{n}}(\vec{q})&\equiv \frac{i}{\hbar }\Bigl[(-i)^{\left|\vec{m}\right|}\bm{B}_{\vec{m}}(\vec{q})\bm{c}_{\vec{n}}(\vec{q})\\
      &\quad \quad \quad -\bm{c}_{\vec{n}}(\vec{q})(+i)^{\left|\vec{m}\right|}\bm{B}_{\vec{m}}(\vec{q})^{\dagger }\Bigr],
     \end{aligned}
  \end{align}
\end{subequations}
with the auxiliary matrices
\begin{subequations}
  \begin{align}
    &\bm{A}_{s}(\vec{q})\equiv \frac{\varepsilon }{\sqrt {\mathstrut 2}\alpha _{s}}\frac{\partial \bm{U}_{0}(\vec{q})}{\partial q_{s}}
    \intertext{and}
    &\begin{aligned}
       \bm{B}_{\vec{m}}(\vec{q})&\equiv \frac{1}{\sqrt {2}^{\left|\vec{m}\right|}\vec{\alpha }\,^{\vec{m}}\vec{m}\,!}\\
       &\quad \quad \times e^{+\varepsilon \beta \bm{U}_{0}(\vec{q})/2}\frac{\partial ^{\left|\vec{m}\right|}\bm{U}\left(\vec{q}\right)}{\partial \vec{q}\,^{\vec{m}}}e^{-\varepsilon \beta \bm{U}_{0}(\vec{q})/2}.
     \end{aligned}
  \end{align}
\end{subequations}
Details regarding the derivation of Eq.~\eqref{eq:brinkman-rep} are presented in Supplemental Information~\nolink{\ref{sec:s-brinkman-derivation}}.

We note that Eq.~\eqref{eq:brinkman-rep} has been employed as the equations of motion for the Brinkman hierarchy to describe a classical Fokker-Planck equation \cite{risken1989book, ikeda2015jcp, ito2016jcp}.
The first and second terms in Eq.~\eqref{eq:brinkman-rep} are the kinetic and potential terms.
The potential term involves the higher-order Moyal expansion terms.
The last term in Eq.~\eqref{eq:brinkman-rep} describes the fluctuation and dissipation that arise from the heat baths.
While $\varepsilon $ is an arbitrary constant, we found that the best choice to enhance the numerical stability of Eq.~\eqref{eq:brinkman-rep} is $\varepsilon =1/2$, as in a case of the single PES \cite{risken1989book}.
Note that the Brinkman hierarchy representation is suitable for the case that the force described by the PESs is not strong, i.e.~the case in which wavepackets in the momentum space is close to the thermal equilibrium state and the wavepackets can be described by a small number of coefficients $\bm{c}_{\vec{n}}(\vec{q},t)$.

\section{Models}
\label{sec:models}
\subsection{Potential energy surfaces and diabatic coupling}
We consider an IC system with two diabatic states, $|0\rangle $ and $|1\rangle $, the reactant and product states, respectively.
These states are strongly coupled to two reaction coordinates, $q_{\mathrm{t}}$ and $q_{\mathrm{c}}$, corresponding to the tuning and coupling modes, respectively.
The tuning mode tunes the energy gap between two electronic states, while the coupling mode couples two electronic diabatic state in the CI case \cite{koppel1984acp, domcke1997acp}.
The diabatic PESs are described by the following combinations of the Morse and harmonic potential curves:
\begin{subequations}
  \begin{align}
    &\begin{aligned}
       U^{\mathrm{00}}(\vec{q})&=D^{0}\left(1-e^{-\sqrt {\hbar \mathstrut \omega _{\mathrm{t}}/2D^{0}}(q_{\mathrm{t}}-q_{\mathrm{t}}^{0})}\right)^{2}\\
       &\quad \quad +\frac{\hbar \omega _{\mathrm{c}}}{2}(q_{\mathrm{c}}-q_{\mathrm{c}}^{0})^{2}+E^{0}\\
       \label{eq:u00}
     \end{aligned}
    \intertext{and}
    &\begin{aligned}
       U^{\mathrm{11}}(\vec{q})&=D^{1}\left(1-e^{+\sqrt {\hbar \mathstrut \omega _{\mathrm{t}}/2D^{1}}(q_{\mathrm{t}}-q_{\mathrm{t}}^{1})}\right)^{2}\\
       &\quad \quad +\frac{\hbar \omega _{\mathrm{c}}}{2}(q_{\mathrm{c}}-q_{\mathrm{c}}^{1})^{2}+E^{1},
       \label{eq:u11}
     \end{aligned}
  \end{align}  
\end{subequations}
where $D^{j}$ is the dissociation energy, $E^{j}$, $q_{\mathrm{t}}^{j}$, and $q_{\mathrm{c}}^{j}$ are the equilibrium minima and displacements of the diabatic PES for $j=0\text{\ and\ }1$, respectively, and $\omega _{\mathrm{t}}$ and $\omega _{\mathrm{c}}$ are the vibrational frequencies at the minimum of the PESs in the $q_{\mathrm{t}}$ and $q_{\mathrm{c}}$ directions.
Note that when a molecule we consider obeys a certain point-group symmetry, we can clearly separate the vibrational modes into the tuning and coupling modes \cite{koppel1984acp, domcke1997acp}.
In our model, this condition is expressed as $q_{\mathrm{c}}^{0}=q_{\mathrm{c}}^{1}$.
When the molecule is non-symmetric, however, the mixing of the tuning and coupling modes occurs.
In our model, this condition is expressed as $q_{\mathrm{c}}^{0}\neq q_{\mathrm{c}}^{1}$, which implies that the coupling mode $q_{\mathrm{c}}$ also tunes the energy gap, and the position of the stable points in the $q_{\mathrm{c}}$ direction change during the reaction processes.

For the case $q_{\mathrm{t}}^{0}<q_{\mathrm{t}}^{1}$, the crossing curve of the diabatic PESs is defined as $\vec{q}\,^{\ast }\equiv (q_{\mathrm{t}}^{\ast },q_{\mathrm{c}}^{\ast })$, where $q_{\mathrm{t}}^{\ast }$ and $ q_{\mathrm{c}}^{\ast }$ satisfy $ U^{\mathrm{00}}({\vec{q}\,}^{\ast })=U^{\mathrm{11}}({\vec{q}\,}^{\ast })$ and $q_{\mathrm{t}}^{0}<q_{\mathrm{t}}^{\ast }<q_{\mathrm{t}}^{1}$.
We chose the parameter values of the PESs such that this crossing curve exists.
Then, we define the reactant and product regions, $\mathcal{R}$ and $\mathcal{P}$, as $q_{\mathrm{t}}\leq q_{\mathrm{t}}^{\ast }$ and $q_{\mathrm{t}}>q_{\mathrm{t}}^{\ast }$, respectively, under the condition $q_{\mathrm{c}}=q_{\mathrm{c}}^{\ast }$.

We introduce the diabatic coupling function as $U^{\mathrm{10}}(\vec{q})=U^{\mathrm{01}}(\vec{q})\equiv \sigma (\vec{q})$, and consider the CI and AC models in which
\begin{subequations}
  \begin{align}
    \sigma (\vec{q})&=c(q_{\mathrm{c}}-q_{\mathrm{c}}^{\mathrm{CI}})
    \label{eq:ci-models}
    \intertext{and}
    \sigma (\vec{q})&=a
    \label{eq:ac-models},
  \end{align}  
\end{subequations}
where $c\neq 0$ and $a\neq 0$ are the coupling constants, and $q_{\mathrm{c}}^{\mathrm{CI}}$ is the CI point in the $q_{\mathrm{c}}$ direction.
Here, the adiabatic PESs cross at a point in the CI case, whereas they do not cross in the AC case, while the diabatic PESs cross in both CI and AC cases.

In terms of $U^{00}(\vec{q})$, $U^{11}(\vec{q})$, $U^{01}(\vec{q})$, and $U^{10}(\vec{q})$, the adiabatic BO PESs on the adiabatic ground and excited states, $|\Phi ^{g}(\vec{q})\rangle $ and $|\Phi ^{e}(\vec{q})\rangle $, are given by
\begin{subequations}
  \begin{align}
    U_{\mathrm{a}}^{g}(\vec{q})&=U^{11}(\vec{q})-\chi (\vec{q})
    \label{eq:ground}
    \intertext{and}
    U_{\mathrm{a}}^{e}(\vec{q})&=U^{00}(\vec{q})+\chi (\vec{q}),
    \label{eq:excited}
  \end{align}
\end{subequations}
where
\begin{align}
  \begin{split}
    \chi (\vec{q})&\equiv \frac{U^{\mathrm{11}}(\vec{q})-U^{\mathrm{00}}(\vec{q})}{2}\\
    &\quad +\frac{1}{2}\sqrt {\bigl(U^{\mathrm{11}}(\vec{q})-U^{\mathrm{00}}(\vec{q})\bigr)^{2}+4{U^{\mathrm{10}}(\vec{q})}^{2}\mathstrut }.
    \label{eq:chi}
  \end{split}
\end{align}
In this case, the transformation matrix given in Eq.~\eqref{eq:transformation-matrix} is expressed as
\begin{align}
  \begin{pmatrix}
    Z^{0g}(\vec{q})\!\!\! & Z^{0e}(\vec{q})\\
    Z^{1g}(\vec{q})\!\!\! & Z^{1e}(\vec{q})\\
  \end{pmatrix}
  &=\frac{1}{\sqrt {\chi (\vec{q})^{2}+\mathstrut \sigma (\vec{q})^{2}}}
  \begin{pmatrix}
    \phantom{+}\chi (\vec{q})\!\!\! & {+}\sigma (\vec{q})\\
    {-}\sigma (\vec{q})\!\!\! & \phantom{+}\chi (\vec{q})\\
  \end{pmatrix}
  \label{eq:transformation-matrix-2state}.
\end{align}
Details of Eqs.~\eqref{eq:ground}, \eqref{eq:excited}, \eqref{eq:chi}, and \eqref{eq:transformation-matrix-2state} are presented in Supplemental Information~\nolink{\ref{sec:s-adiabatic}}.
The minimum point of the adiabatic excited PES, $\vec{q}\,^{e,\mathrm{min}} \equiv (q_{\mathrm{t}}^{e,\mathrm{min}},q_{\mathrm{c}}^{e,\mathrm{min}})$, is determined as the minimum point of Eq.~\eqref{eq:excited}.

\subsection{Distributions}
We define the expectation value of the kinetic energy for the mode $q_{s}$ as
\begin{align}
  K_{s}(t)&\equiv \int \!d\vec{p}\,\int \!d\vec{q}\,\frac{\hbar \omega _{s}p_{s}^{2}}{2}\sum _{a}W_{\mathrm{a}}^{aa}(\vec{p},\vec{q},t).
\end{align}
We also define the ground and excited state distributions, $W^{g}(\vec{p},\vec{q},t)\equiv W_{\mathrm{a}}^{gg}(\vec{p},\vec{q},t)$ and $W^{e}(\vec{p},\vec{q},t)\equiv W_{\mathrm{a}}^{ee}(\vec{p},\vec{q},t)$.
The ground state distribution is further divided into the reactant and product states, $W^{0}(\vec{p},\vec{q},t)$ and $W^{1}(\vec{r},\vec{q},t)$, for $\vec{q}$ in the regions $\mathcal{R}$ and $\mathcal{P}$, respectively.
Accordingly, we introduce the probability distribution and population for $\phi =e$, $g$, $0$, and $1$ as
\begin{align}
  F^{\phi }(\vec{q},t)&\equiv \int d\vec{p}\,W^{\phi }(\vec{p},\vec{q},t),
\end{align}
and
\begin{align}
  u^{\phi }(t)&\equiv \int d\vec{q}\,F^{\phi }(\vec{q},t),
\end{align}
where $u^{g}(t)=u^{0}(t)+u^{1}(t)$ and $u^{g}(t)+u^{e}(t)=1$.
Then the expectation value of $q_{s}$ at time $t$ is defined as
\begin{align}
  \langle q_{s}(t)\rangle ^{\phi }&\equiv \int d\vec{q}\,q_{s}F^{\phi }(\vec{q},t)/u^{\phi }(t).
\end{align}
The Brinkman hierarchy representation of these conditions and expectation values are presented in Supplemental Information~\nolink{\ref{sec:s-c-misc}}.

\section{Numerical Results}
\label{sec:results}
\begin{table}
  \centering
  \caption{ The parameter values for the CI and AC models. }
  \begin{tabular}{cc|c@{~}r@{~}l}
    \hline
    \hline
    \multicolumn{2}{c|}{Model} &
    Symbol & \multicolumn{2}{c}{Value}\\
    \hline
    \hline
    \multicolumn{2}{c|}{\multirow{12}{*}{--}} 
    & $\omega _{\mathrm{t}}$ & $100$&$\mathrm{cm}^{-1}$\\
    && $\omega _{\mathrm{c}}$ & $500$&$\mathrm{cm}^{-1}$\\
    && $D^{0}$  & $20,000$&$\mathrm{cm}^{-1}$\\
    && $D^{1}$ & $20,000$&$\mathrm{cm}^{-1}$\\
    && $E^{0}$  & $0$&$\mathrm{cm}^{-1}$\\
    && $E^{1}$ & $2,000$&$\mathrm{cm}^{-1}$\\
    && $q_{\mathrm{t}}^{0}$ & $-10$&\\
    && $q_{\mathrm{t}}^{1}$ & $+10$&\\
    && $\zeta _{\mathrm{t}}$ & $100$&$\mathrm{cm}^{-1}$ \\
    && $\zeta _{\mathrm{c}}$ & $50$&$\mathrm{cm}^{-1}$ \\
    && $\beta $ & $1/208.51$&$\mathrm{cm}$ \\
    && ($T$ & $300$&$\mathrm{K}$) \\
    \cline{2-5}
    \multirow{6}{*}{CI} && $c$ & $200$&$\mathrm{cm}^{-1}$ \\
    && $d$ & $-2.5$ --- $+2.5$ & \\
    \cline{3-5}
    &\multirow{2}{*}{0}& $q_{\mathrm{c}}^{0}$ & $0$  & \\
    & & $q_{\mathrm{c}}^{1}$ & $0$  & \\
    \cdashline{3-5}
    &\multirow{2}{*}{1}& $q_{\mathrm{c}}^{0}$ & $-0.5$  & \\
    &  & $q_{\mathrm{c}}^{1}$ & $+0.5$ & \\
    \cline{2-5}
    \multirow{5}{*}{AC}
    && $a$ & $50$ --- $500$&$\mathrm{cm}^{-1}$ \\
    \cline{3-5}
    &\multirow{2}{*}{0}& $q_{\mathrm{c}}^{0}$ & $0$ \\
    & & $q_{\mathrm{c}}^{1}$ & $0$  &  \\
    \cdashline{3-5}
    &\multirow{2}{*}{1}& $q_{\mathrm{c}}^{0}$ & $-0.5$ \\
    & & $q_{\mathrm{c}}^{1}$ & $+0.5$  \\
    \hline
    \hline
  \end{tabular}
  \label{tab:parameters}
\end{table}
We carried out numerical simulations for the system described by the diabatic PESs, Eqs.~\eqref{eq:u00} and \eqref{eq:u11}.
Next, we introduce the CI and AC models described by the diabatic coupling given in Eqs.~\eqref{eq:ci-models} and \eqref{eq:ac-models}.
For each model, we consider the case with and without displacements in the coupling mode $q_{\mathrm{c}}$.
Thus, we have four models, which we call CI0, CI1, AC0, and AC1, where $0$ and $1$ correspond to the cases $q_{\mathrm{c}}^{1}-q_{\mathrm{c}}^{0}=0$ and $q_{\mathrm{c}}^{1}-q_{\mathrm{c}}^{0}=1$.
For $q_{\mathrm{c}}^{1}-q_{\mathrm{c}}^{0}\neq 0$, the coupling mode also tunes the energy gap.
To investigate the efficiency of the IC process, we introduce the CI-minimum distance defined as
\begin{align}
  d&\equiv q_{\mathrm{c}}^{\mathrm{CI}}-q_{\mathrm{c}}^{e,\mathrm{min}},
\end{align}
where $q_{\mathrm{c}}^{\mathrm{CI}}$ and $q_{\mathrm{c}}^{e,\mathrm{min}}$ are the CI point and the minimum of the adiabatic excited PES.
The parameter values for the system and the heat baths are listed in Table~\ref{tab:parameters}.

Numerical calculations were carried out to integrate Eq.~\eqref{eq:brinkman-rep} using the fourth-order low-storage Runge-Kutta method with a time step $\delta t=0.2~\mathrm{fs}$ \cite{williamson1980jcp,yan2017cjcp}.
The finite difference calculations for the $\vec{q}$ derivatives in Eq.~\eqref{eq:brinkman-rep} were performed using the first-order difference method with ninth-order accuracy \cite{fornberg1988mc}.
The numerical integrations of the MSWDF in $\vec{q}$ space were performed using the trapezoidal rule.
The mesh size in $\vec{q}$ space was $N_{\mathrm{t}}\times N_{\mathrm{c}}=128\times 48$ with mesh ranges $-20\leq q_{\mathrm{t}}\leq +20$ and $-8\leq q_{\mathrm{c}}\leq +8$.
The depth of the Brinkman hierarchy was $40$ and $32$ for $n_{\mathrm{t}}$ and $n_{\mathrm{c}}$, respectively.
We introduced artificial viscosity terms in Eq.~\eqref{eq:brinkman-rep} to enhance the stability of the numerical integrations.
The details regarding the finite differences and artificial viscosity terms are explained in Supplemental Information~\nolink{\ref{sec:s-artificial-viscosity}}.

It has been shown that a GPGPU, which utilizes fast access memory and super parallel architecture, is a powerful device to integrate simultaneous differential equations for systems with many degrees of freedom, as Eq.~\eqref{eq:brinkman-rep}.
We carried out the time integration of Eq.~\eqref{eq:brinkman-rep} using the first-order form of Eq.~\eqref{eq:qle-Liouvillian} given in Eq.~\eqref{eq:pot-qcle} for $1,000~\mathrm{fs}$ employing C++/CUDA codes with an NVIDIA Tesla P100 and employing single thread C++ codes with an Intel Xeon CPU E5-2667.
The computation times were approximately $15$ minutes for the former and $16$ hours for the latter.

\subsection{Verification of the truncated Moyal expansion}
\label{sec:truncation}
We introduce the truncated Moyal product defined as
\begin{align}
  \star _{(M)}&\equiv \sum _{m=0}^{M}\frac{1}{m!}\left[{\sum _{s}\frac{i}{2}\Bigl(\underaccent{\leftarrow }{\partial }_{q_{s}}\underaccent{\rightarrow }{\partial }_{p_{s}}-\underaccent{\rightarrow }{\partial }_{q_{s}}\underaccent{\leftarrow }{\partial }_{p_{s}}\Bigr)}\right]^{m},
  \label{eq:truncated-moyal-star}
\end{align}
where $M$ is the order of the truncation.
Then Eqs.~\eqref{eq:qle-Liouvillian} and \eqref{eq:adiabatic-transform} can be written as
\begin{subequations}
  \begin{align}
    \begin{split}
      &\mathcal{L}_{W}(\vec{p},\vec{q})\bm{W}(\vec{p},\vec{q})\\
      &=\sum _{s}\omega _{s}p_{s}\frac{\partial }{\partial q_{s}}\bm{W}\left(\vec{p},\vec{q}\right)\\
      &\quad \quad +\frac{i}{\hbar }\Bigl[\bm{U}\left(\vec{q}\right)\star _{(\mathcal{M})}\bm{W}\left(\vec{p},\vec{q}\right)-\bm{W}\left(\vec{p},\vec{q}\right)\star _{(\mathcal{M})}\bm{U}\left(\vec{q}\right)\Bigr],
    \end{split}
    \intertext{and}
    \begin{split}
      &\bm{W}_{\mathrm{a}}(\vec{p},\vec{q},t)\\
      &=\sum _{M+M'=\mathcal{M}-1}\bm{Z}\left(\vec{q}\right)^{\dagger }\star _{(M)}\bm{W}(\vec{p},\vec{q},t)\star _{(M')}\bm{Z}\left(\vec{q}\right),
    \end{split}
  \end{align}  
\end{subequations}
respectively, where $\mathcal{M}\geq 1$.
Before carrying out the full analysis, we examine the accuracy of the above expressions.
For this purpose, we consider the AC0 case, in which the coupling mode, $q_{\mathrm{c}}$, decouples from the tuning mode, $q_{\mathrm{t}}$, and the electronic states.
For this reason, we can ignore the dynamics in the $q_{\mathrm{c}}$ direction and regard the AC0 model as a one-dimensional system.

To obtain a proper initial distribution that is consistent with the truncation Eq.~\eqref{eq:truncated-moyal-star} and the high-temperature approximation, we first set the initial distribution as the classical Boltzmann distribution of the reactant adiabatic ground state, which is expressed as $W_{\mathrm{a}}^{ee}(p_{\mathrm{t}},q_{\mathrm{t}},-t_{\mathrm{eq}})=W_{\mathrm{a}}^{eg}(p_{\mathrm{t}},q_{\mathrm{t}},-t_{\mathrm{eq}})=W_{\mathrm{a}}^{ge}(p_{\mathrm{t}},q_{\mathrm{t}},-t_{\mathrm{eq}})=0$, $W_{\mathrm{a}}^{gg}(p_{\mathrm{t}},q_{\mathrm{t}},-t_{\mathrm{eq}})=e^{-\beta \hbar \omega _{\mathrm{t}}p_{\mathrm{t}}^{2}/2}e^{-\beta U_{\mathrm{a}}^{g}(q_{\mathrm{t}})}/\mathcal{Z}$ for $q_{\mathrm{t}}$ in the region $\mathcal{R}$, and as $\allowbreak W_{\mathrm{a}}^{gg}(p_{\mathrm{t}},q_{\mathrm{t}},-t_{\mathrm{eq}})=0$ otherwise, where $\mathcal{Z}$ is the partition function.
Then we integrate Eq.~\eqref{eq:qfpe} for a sufficiently long time $t_{\mathrm{eq}}$ to obtain a stationary solution, expressed as $\bar{\bm{W}}(p_{\mathrm{t}},q_{\mathrm{t}})$.

We assume that the reaction process is initiated by photo-excitation created by a pair of impulsive pulses at $t=0$ described by the dipole moment $\{\bm{\mu }\}^{jk}\equiv \langle j|\hat{\mu }|k\rangle $ for $j,k=0,1$ as
\begin{align}
  \begin{split}
    \bm{W}(p_{\mathrm{t}},q_{\mathrm{t}},0)&=\bar{\bm{W}}(p_{\mathrm{t}},q_{\mathrm{t}})\\
    &\quad \quad +(\bar{E}\Delta \tau )^{2}\frac{i}{\hbar }\biggl[\bm{\mu },\frac{i}{\hbar }\Bigl[\bm{\mu },\bar{\bm{W}}(p_{\mathrm{t}},q_{\mathrm{t}})\Bigr]\biggr],
  \end{split}
  \label{eq:excitation}
\end{align}
where $\bar{E}$ and $\Delta \tau $ are the electric field amplitude and pulse duration of the two pulses, and the square brackets represent the ordinary commutator, defined as $[\bm{A},\bm{B}]\equiv \bm{A}\bm{B}-\bm{B}\bm{A}$.
Equation~\eqref{eq:excitation} consists of the zeroth-order and second-order perturbative expansion terms of the electric field, and the excited wavepacket dynamics measured in impulsive pump-probe spectroscopy can be simulated from the second-order term \cite{mukamel1999book}.
For the elements $\bar{W}^{jk}\equiv \bar{W}^{jk}(p_{\mathrm{t}},q_{\mathrm{t}})$ with $\hat{\mu }=\mu (|0\rangle \langle 1|+|1\rangle \langle 0|)$, $\bm{W}(p_{\mathrm{t}},q_{\mathrm{t}},0)$ is given in matrix form as
\begin{align}
  \begin{split}
    &\bm{W}(p_{\mathrm{t}},q_{\mathrm{t}},0)\\
    &=
    \begin{pmatrix}
      \bar{W}^{00}+A\left( \bar{W}^{11}-\bar{W}^{00}\right)
      &\bar{W}^{01}+A\left(\bar{W}^{10}-\bar{W}^{01} \right)\\ \bar{W}^{10}+A\left(\bar{W}^{01}-\bar{W}^{10} \right)
      &\bar{W}^{11}+A\left( \bar{W}^{00}-\bar{W}^{11}\right)
   \end{pmatrix},
  \end{split}
\end{align}
where $A \equiv 2(\bar{E}\mu \Delta \tau )^{2}/\hbar ^{2}$.
Thus, when we choose the dimensionless constant $A$ to be $A=1$, the excitation pulses cause the populations of $|0\rangle $ and $|1\rangle $ states to be exchanged.

\begin{figure}
  \centering
  \includegraphics[scale=0.9]{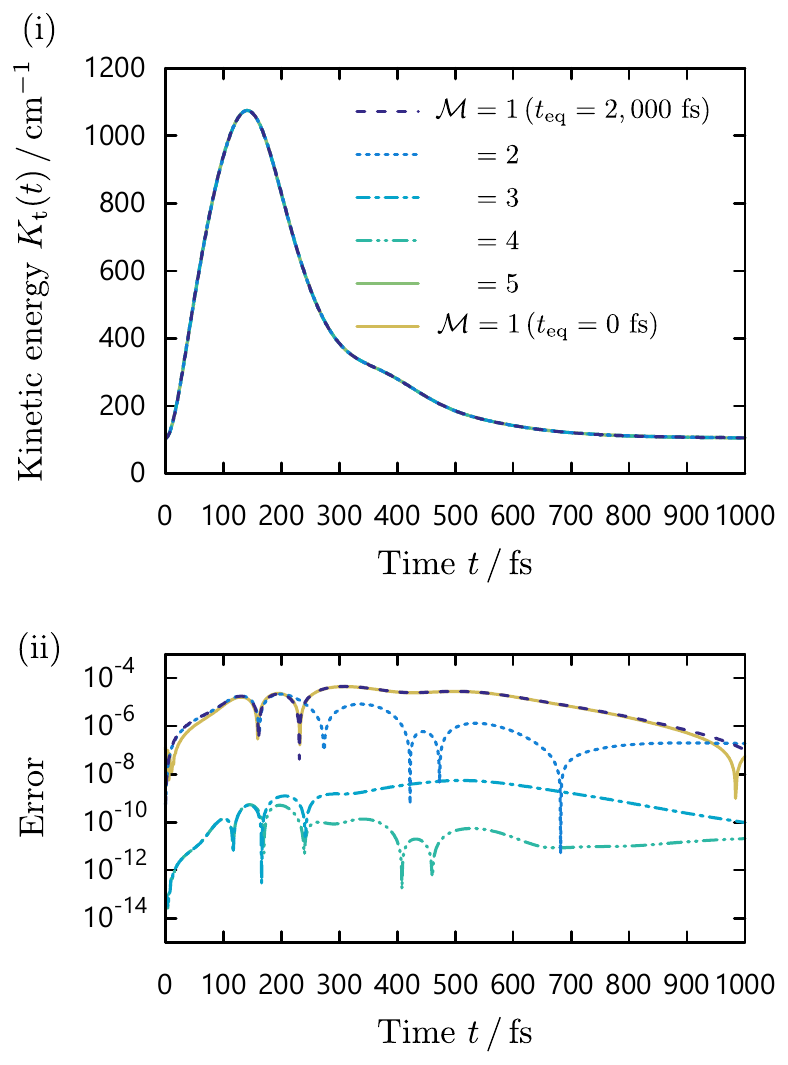}
  \caption{ (i) The kinetic energy of the tuning mode $K_{\mathrm{t}}(t)$ for the AC0 model is plotted as a function of time for various values of the truncation depth, $\mathcal{M}=1$---$5$, with the initial equilibration time $t_{\mathrm{eq}}=2,000~\mathrm{fs}$.
    For the case $\mathcal{M}=1$, we obtained the same result without the initial equilibration, i.e.~using $t_{\mathrm{eq}}=0~\mathrm{fs}$.
    All of the calculated curves are nearly coincident and cannot be distinguished in this graph.
    (ii) The relative errors defined as $\left|K_{\mathrm{t}}(t)-K_{\mathrm{t}}^{\mathcal{M}=5}(t)\right|/\left|\max _{t'}\left\{K_{\mathrm{t}}^{\mathcal{M}=5}(t')\right\}\right|$ for the calculations plotted in (i). }
  \label{fig:truncation}
\end{figure}
In Fig.~\ref{fig:truncation}, we depict the kinetic energy of the tuning mode $K_{\mathrm{t}}(t)$ for fixed $a=200~\mathrm{cm}^{-1}$ from $M=1$ to 5.
As seen there, the results obtained with the first-order truncation $\mathcal{M}=1$ are very similar to those from $\mathcal{M}=5$.
This is because the damping operators in the MSFPE cause the profile of the wavepackets to be Gaussian-like and because the anharmonicity of the Morse potential is not strong.

Here, we have examined the numerical accuracy for the AC0 case, but the same argument can be applied to the CI0, CI1, and AC1 case, because the diabatic PESs in the $q_{\mathrm{c}}$ direction are harmonic, and as a result the truncation for $\mathcal{M}=1$ is analytically exact.
Hence, hereafter we employ $\mathcal{M}=1$.

\subsection{Wavepacket dynamics}
\label{sec:wavepackets}
In the previous subsection, we demonstrated the accuracy of the numerical calculations starting from the true thermal equilibrium sate.
For the case $\mathcal{M}=1$, however, the difference between the classical Boltzmann equilibrium state and the true thermal equilibrium state described by the MSQFPE is minor, as shown in the case for $t_{\mathrm{eq}}=0~\mathrm{fs}$ in Fig.~\ref{fig:truncation}.
For this reason, we use the classical Boltzmann distribution of the adiabatic ground state as the initial state.
We thus set the initial distribution for $\vec{q}$ in the region $\mathcal{R}$ as
\begin{align}
  W_{\mathrm{a}}^{ee}(\vec{p},\vec{q},0)=\frac{1}{\mathcal{Z}}
  e^{-\beta \left(\hbar \omega _{\mathrm{t}}p_{\mathrm{t}}^{2}/2+\hbar \omega _{\mathrm{c}}p_{\mathrm{c}}^{2}/2+U_{\mathrm{a}}^{g}(\vec{q})\right)},
  \label{eq:initial-2d}
\end{align}
and as $\bm{W}_{\mathrm{a}}(\vec{p},\vec{q},0)=0$ otherwise. 

\begin{figure}
  \centering
  \includegraphics[scale=0.9]{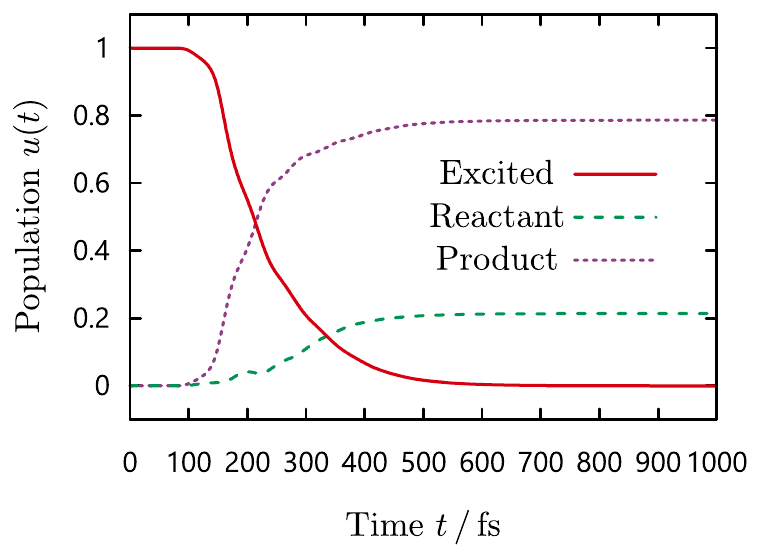}
  \caption{ The time evolution of the population in the CI1 model with $d=1.5$.
    The red, green, and purple curves represent the populations for the excited, ground reactant, and ground product states, $u^{e}(t)$, $u^{0}(t)$, and $u^{1}(t)$, respectively.
    Here, the lifetime of the excited population, given in Eq.~\eqref{eq:lifetime}, and the yield, given in Eq.~\eqref{eq:product-ratio}, are evaluated as $\tau ^{e}=235~\mathrm{fs}$ and $y^{\mathrm{p}}=0.788$, respectively. }
  \label{fig:population}
\end{figure}
Figure~\ref{fig:population} illustrates the time evolution of the populations in the excited, ground reactant, and ground product states, $u^{e}(t)$, $u^{0}(t)$, and $u^{1}(t)$, with the CI-minimum distance $d=1.5$.
For the excited state, the excited wavepacket is initially in the reactant region on the excited state, and then the wavepacket arrives at the crossing region at approximately $t=100~\mathrm{fs}$.
Because the excited wavepacket has kinetic energy obtained from the potential in the direction of the product region at the CI point, most of the population quickly transfers to the product region in the ground state.

\begin{figure}
  \centering
  \includegraphics[scale=0.9]{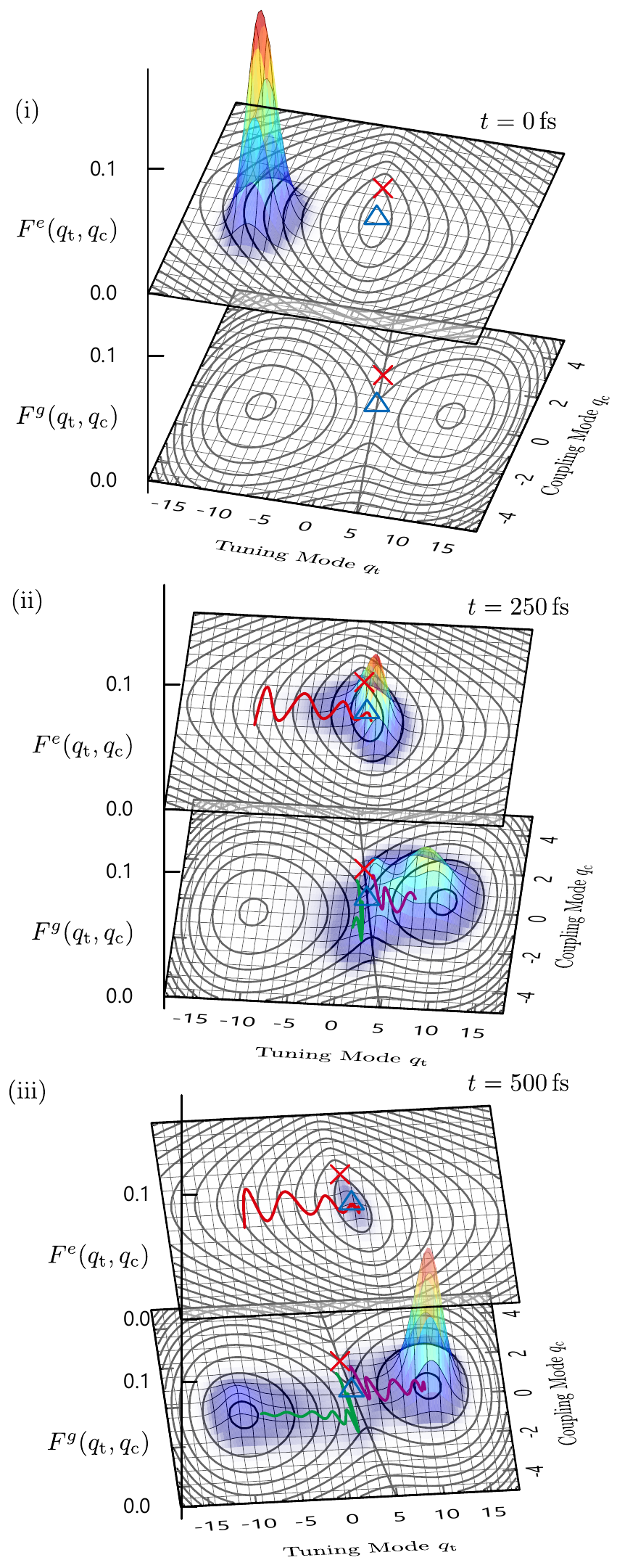}
  \caption{ Snapshots of excited wavepackets in CI1 model with waiting times (i) $t=0~\mathrm{fs}$, (ii) $250~\mathrm{fs}$, and (iii) $500~\mathrm{fs}$.
    The CI-minimum distance is $d=1.5$.
    The red, green, and purple curves represent the mean coordinates for the excited, ground reactant, and ground product states, $\langle \vec{q}(t)\rangle ^{e}$, $\langle \vec{q}(t)\rangle ^{0}$, and $\langle \vec{q}(t)\rangle ^{1}$, respectively.
    The contours of the adiabatic BO PESs for every $1,000~\mathrm{cm}^{-1}$ are represented by the black curves in the two-dimensional plates.
    The symbols $\times $ and $\triangle $ represent the CI point and the minimum of $U_{\mathrm{a}}^{e}(\vec{q})$, respectively.
    Videos of excited wavepackets in CI1($d=+1.5$), CI1($d=-1.5$) and AC1($a=300~\mathrm{cm}^{-1}$) models are presented in Supplemental Information.
  }
  \label{fig:fig4-wavepackets}
\end{figure}
For the time evolution of $u^{0}(t)$, we observe slight oscillatory behavior at approximately $t=200~\mathrm{fs}$, when the wavepackets are in the crossing region.
To analyze this point more closely, we display snapshots of wavepackets in the ground and excited adiabatic states, $F^{g}(\vec{p},\vec{q},t)$ and $F^{e}(\vec{p},\vec{q},t)$, in Fig.~\ref{fig:fig4-wavepackets}.
Due to the difference between the stable points for the adiabatic ground and excited states in the direction of $q_{\mathrm{c}}$, the wavepacket exhibits vibrational motion in the direction of the coupling mode, $q_{\mathrm{c}}$.
As depicted in Fig.~\ref{fig:fig4-wavepackets}(ii), the wavepacket cannot be centered at the CI point because of this vibrational motion.
As a result, the population of the reactant also exhibits oscillatory motion, as observed in Fig.~\ref{fig:population}.
Note that the vibrational motions in the coupling mode induce non-adiabatic transitions for the case $q_{\mathrm{c}}^{0}\neq q_{\mathrm{c}}^{1}$, because the coupling mode also couples to the electronic states.
After the oscillatory motion of the excited state wavepacket in the $q_{\mathrm{c}}$ direction and the motion in the $q_{\mathrm{t}}$ direction relax, a small portion of the wavepacket is trapped at the local minimum in the crossing region.
This wavepacket slowly relaxes to both the reactant and product regions in the ground state.
By $t=500~\mathrm{fs}$, illustrated in Fig.~\ref{fig:fig4-wavepackets}(iii), has excited wavepacket is relaxed and has mostly transferred to the product states.

\subsection{The lifetime of the adiabatic excited state}
\label{sec:lifetime}
\begin{figure}
  \centering
  \includegraphics[scale=0.9]{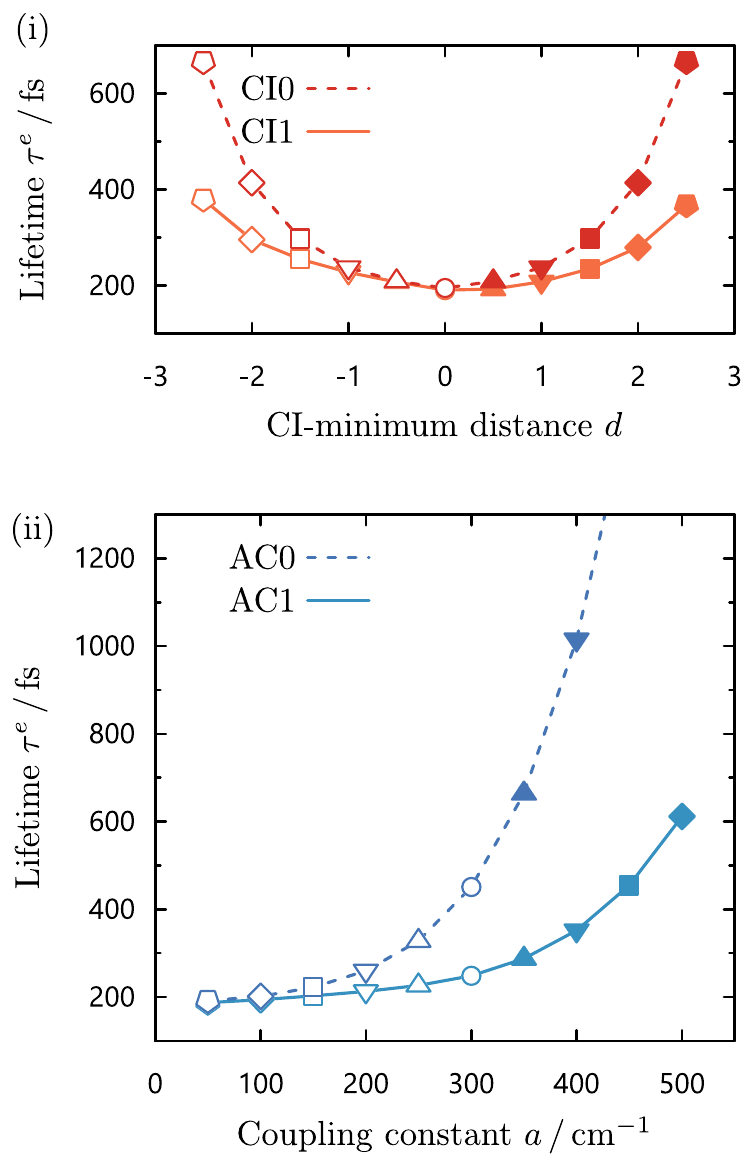}
  \caption{ (i) The lifetime of the adiabatic excited state $\tau ^{e}$ as a function of $d\equiv q_{\mathrm{c}}^{\mathrm{CI}}-q_{\mathrm{c}}^{e,\mathrm{min}}$ in the CI0 (the dashed curve) and CI1 (the solid curve) cases, where $q_{\mathrm{c}}^{\mathrm{CI}}$ and $q_{\mathrm{c}}^{e,\mathrm{min}}$ are the CI point and the minimum of the adiabatic excited PES.
    (ii) The lifetime of the adiabatic excited state $\tau ^{e}$ as a function of the diabatic coupling $a$ in the AC0 (the dashed curve) and AC1 (the solid curve) cases.
  }
  \label{fig:lifetime}
\end{figure}
We next investigate the lifetime of the adiabatic excited state for the CI and AC models with various values of the diabatic coupling parameters.
We define the lifetime $\tau ^{e}$ as
\begin{align}
  \tau ^{e}&\equiv \int _{0}^{t_{\mathrm{f}}}\!dt\,\frac{u^{e}(t)-u^{e}(t_{\mathrm{f}})}{u^{e}(0)-u^{e}(t_{\mathrm{f}})}
  \label{eq:lifetime},
\end{align}
where $t_{\mathrm{f}}$ is a sufficiently long time to evaluate the entire de-excitation dynamics.
In the following calculations, we set $t_{\mathrm{f}}=10,000~\mathrm{fs}$.

Figure \ref{fig:lifetime}(i) plots $\tau ^{e}$ as a function of the CI-minimum distance $d\equiv q_{\mathrm{c}}^{\mathrm{CI}}-q_{\mathrm{c}}^{e,\mathrm{min}}$ for the CI model, where $q_{\mathrm{c}}^{\mathrm{CI}}$ and $q_{\mathrm{c}}^{e,\mathrm{min}}$ are the CI point and the minimum of the adiabatic excited PES.
The lifetime $\tau ^{e}$ is shortest at $d=0$, while it increases with the absolute value of $d$, because the center of the excited wavepackets is away from the CI point for larger $d$.
In the CI1 case, $\tau ^{e}$ is not symmetric as a function of $d$, because the diabatic PESs are not symmetric in the $q_{\mathrm{c}}$ direction for $q_{\mathrm{c}}^{0}\neq q_{\mathrm{c}}^{1}$.
Figure \ref{fig:lifetime}(ii) depicts $\tau ^{e}$ as a function of the diabatic coupling $a$ in the AC model.
Because the energy gap between the adiabatic ground and excited PESs increases as the diabatic coupling increases, the lifetime $\tau ^{e}$ increases monotonically as $a$ increases.
In the case $q_{\mathrm{c}}^{0}\neq q_{\mathrm{c}}^{1}$, the motion of the coupling mode is excited in the $q_{\mathrm{c}}$ direction due to the difference between the stable points of the PESs.
Because of this motion, the BO approximation breaks down.
As a result, the lifetimes are shorter in the CI1 and AC1 cases than in the CI0 and AC0 cases.

\subsection{The yield}
\label{sec:product-ratio}
\begin{figure}
  \centering
  \includegraphics[scale=0.9]{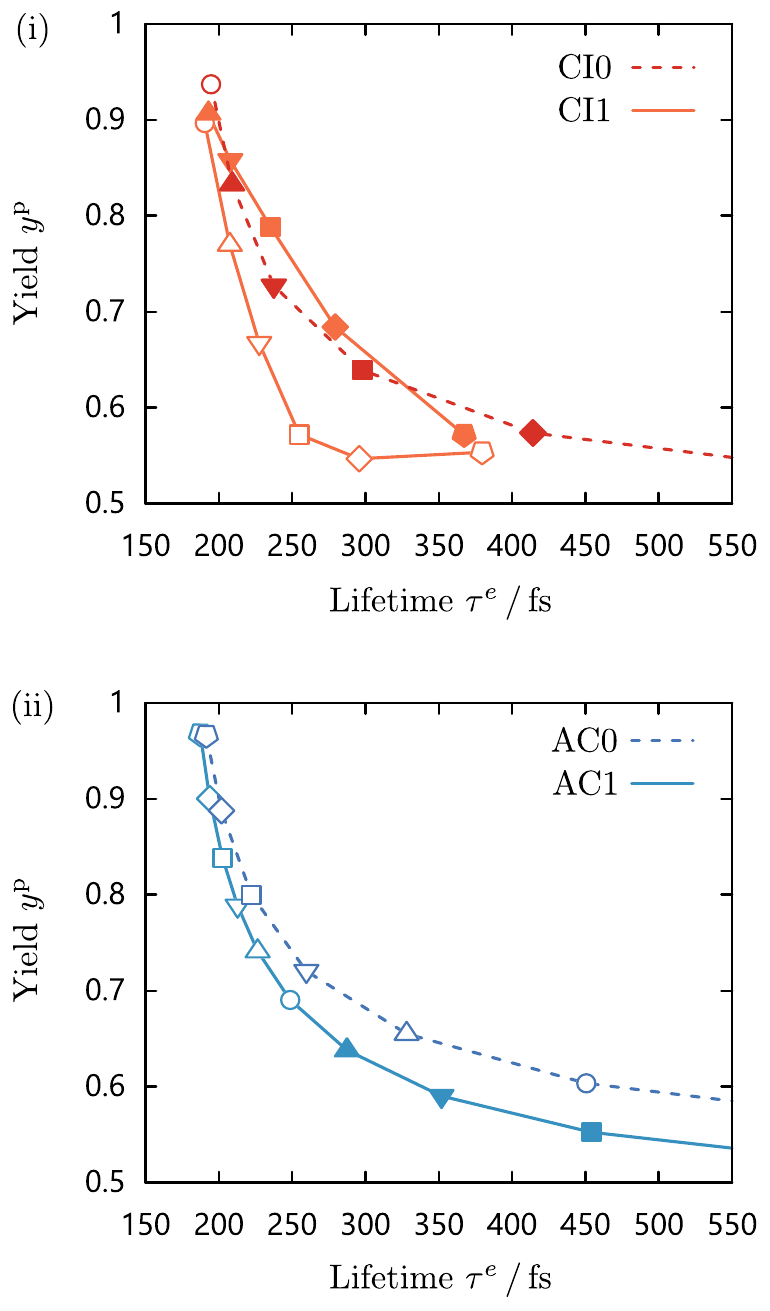}
  \caption{ The yield $y^{\mathrm{p}}$ as a function of the lifetime $\tau ^{e}$ in (i) the CI0 (the dashed curve) and CI1 (the solid curve) cases and (ii) the AC0 (the dashed curve) and AC1 (the solid curve) cases.
    Symbol has the same meaning as in Fig.~\ref{fig:lifetime} for different $d$ and $a$.
  }
  \label{fig:fig-yield}
\end{figure}
In Fig.~\ref{fig:fig-yield}, we illustrate the relationship between the lifetime $\tau ^{e}$ and the yields at the final time, $t_{\mathrm{f}}$, defined as
\begin{align}
  y^{\mathrm{p}}&\equiv u^{1}(t_{\mathrm{f}})/u^{g}(t_{\mathrm{f}}),
  \label{eq:product-ratio}
\end{align}
using the same parameter values as in the calculations considered in Fig.~\ref{fig:lifetime}.

We observe large yields in the CI0, AC0, and AC1 cases, in accordance with their lifetimes $\tau ^{e}$.
Note that a short $\tau ^{e}$ implies that the excited wavepackets move smoothly across the crossing region while maintaining a large momentum in the direction of the product region, as explained in Sec.~\ref{sec:wavepackets}.
The yields $y^{\mathrm{p}}$ in the CI0 and AC1 cases are smaller than in the AC0 case due to the contributions from non-adiabatic transitions in the coupling mode.
The similarity between the CI0 and AC0 cases indicates that when the adiabatic excited PESs are smooth and there is no displacement in the coupling mode, and hence $q_{\mathrm{c}}^{0}=q_{\mathrm{c}}^{1}$, the CI does not play an important role.
In such cases, the results of the IC processes via the CI are described by an AC model that is characterized by an effectively weak diabatic coupling constant, $a$.

Contrastingly, the $\tau ^{e}$-$y^{\mathrm{p}}$ curve in the CI1 case exhibits a bifurcation for negative and positive $d$.
The yields $y^{\mathrm{p}}$ in the case of negative $d$ are smaller than in the other three cases, while the lifetimes $\tau ^{e}$ are short.
This is because, for negative $d$, the excited wavepackets first pass through the crossing region away from the CI point.
Then, after bouncing off of the barrier of the excited PES in the product region and losing energy due to dissipation, the wavepacket encounters the CI point on the way back from the reactant region.
Therefore, although the lifetimes for $d=\pm 1.5$ in the CI1 case are similar, $\tau ^{e}\sim 250~\mathrm{fs}$, the yield in the negative $d$ case is much lower.
This indicates that the photoisomerization process is sensitive to the profiles of the PESs, in particular when the CI plays a significant role.

\subsection{Kinetic energy and momentum distribution of the coupling mode}
\begin{figure}
  \centering
  \includegraphics[scale=0.9]{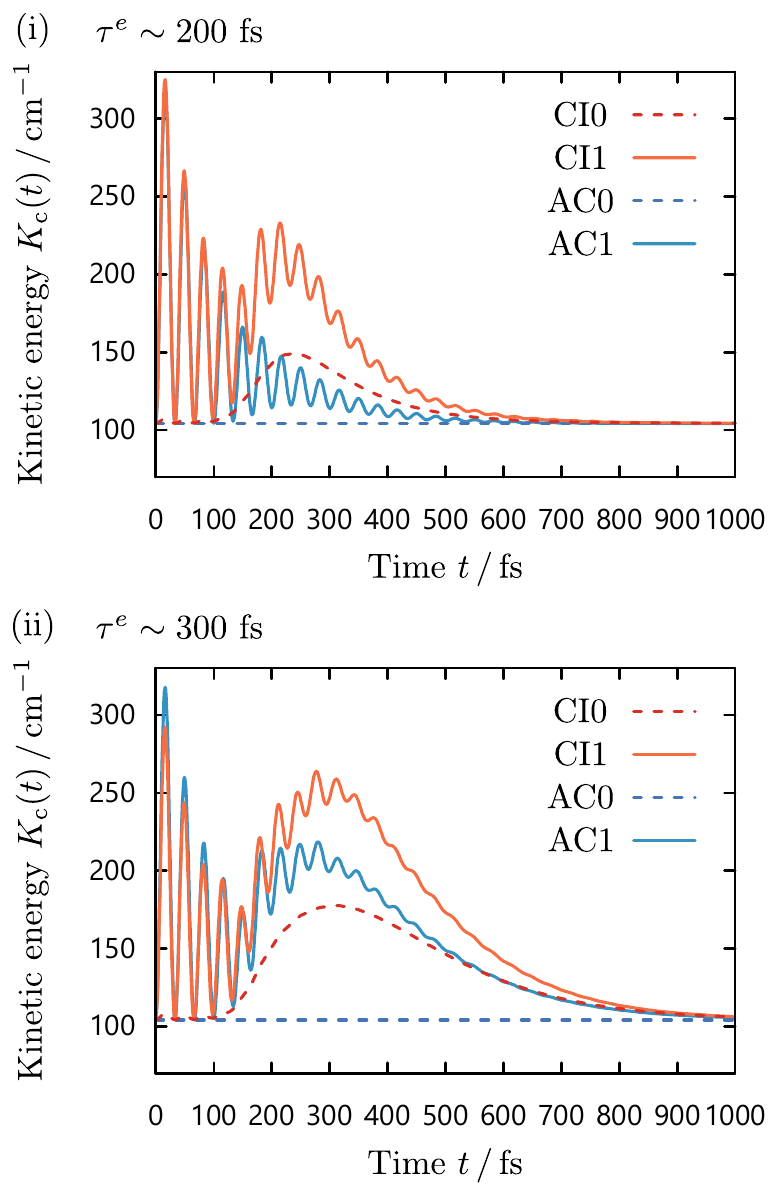}
  \caption{ The kinetic energy of the coupling mode $K_{\mathrm{c}}(t)$ as a function of $t$ for (i) $\tau ^{e}\sim 200~\mathrm{fs}$ and (ii) $\tau ^{e}\sim 300~\mathrm{fs}$ in the CI0 (the red dashed curve), CI1 (the orange solid curve) , AC0 (the blue dashed curve), and AC1 (the cyan solid curve) cases.
    In each case, the lifetime was controlled by adjusting the diabatic coupling parameters.
  }
  \label{fig:K_t}
\end{figure}
\begin{table}
  \centering
  \caption{ Parameters values that we used to investigate the time evolution of $K_{\mathrm{c}}(t)$ and electronic coherence $C^{eg}(\vec{q},t)$.
  }
  \begin{tabular}{c|r@{}cc|r@{}cc}
    \hline
    \hline
    Lifetime $\tau ^{e}$ && CI0 & CI1 && AC0 & AC1 \\
    \hline
    $\sim 200~\mathrm{fs}$ & \multirow{2}{*}{$d=$} & $\phantom{-}0.0$ & $\phantom{-}0.0$ & \multirow{2}{*}{$a/\mathrm{cm}^{-1}=$} & $100$ & $100$ \\
    $\sim 300~\mathrm{fs}$ &      & $-1.5$ & $-2.0$ & & $220$ & $350$ \\
    \hline
    \hline
  \end{tabular}
  \label{tab:parameters-K_t}
\end{table}
In Fig.~\ref{fig:K_t}, we plot the kinetic energy of the coupling mode, $K_{\mathrm{c}}(t)$, as a function of time for the CI and AC models.
In these calculations, we adjusted the diabatic coupling parameters so as to have lifetimes $\tau ^{e}\sim 200~\mathrm{fs}$ and $300~\mathrm{fs}$.
The parameter sets we used are listed in Table \ref{tab:parameters-K_t}.
In the CI0 cases, the excitation energy and the kinetic energy $K_{\mathrm{t}}(t)$ are converted to $K_{\mathrm{c}}(t)$ during the non-adiabatic transition process, and $K_{\mathrm{c}}(t)$ reaches the equilibrium value $\sim k_{\mathrm{B}}T/2=104~\mathrm{cm}^{-1}$ due to thermal relaxation.
In the AC0 cases, in which the coupling mode $q_{\mathrm{c}}$ is independent of the tuning mode $q_{\mathrm{t}}$ and the electronic states, $K_{\mathrm{c}}(t)$ does not change during the non-adiabatic transition process.
In the CI1 and AC1 cases, vibrational motion in the coupling mode arises due to the displacement between the adiabatic ground and excited PESs, $q_{\mathrm{c}}^{1}-q_{\mathrm{c}}^{0}=1$.
Because $K_{\mathrm{c}}(t)$ is proportional to $p_{\mathrm{c}}^{2}$, we observe the vibrational peaks with frequency $2\omega _{\mathrm{c}}=1,000~\mathrm{cm}^{-1}$ that corresponds the period of $33.4~\mathrm{fs}$.
Then the vibrational peaks become larger as a function of time due to the non-adiabatic transitions, and subsequently decreases due to the relaxation induced by the environments.
The maximum value of $K_{\mathrm{c}}(t)$ is large for the long lifetime cases in Fig.~\ref{fig:K_t}(ii), because the wavepackets are trapped in the crossing region and the kinetic energy of the excited state converts to the vibrational energy of the tuning mode, while the wavepackets pass through the crossing region without exciting the coupling mode for the short lifetime cases in Fig.~\ref{fig:K_t}(i). 
These results indicate that, the vibrational dynamics of the coupling mode through the non-adiabatic transition processes are different even when the lifetimes of the excited population are similar.

\begin{figure}
  \centering
  \includegraphics[scale=0.9]{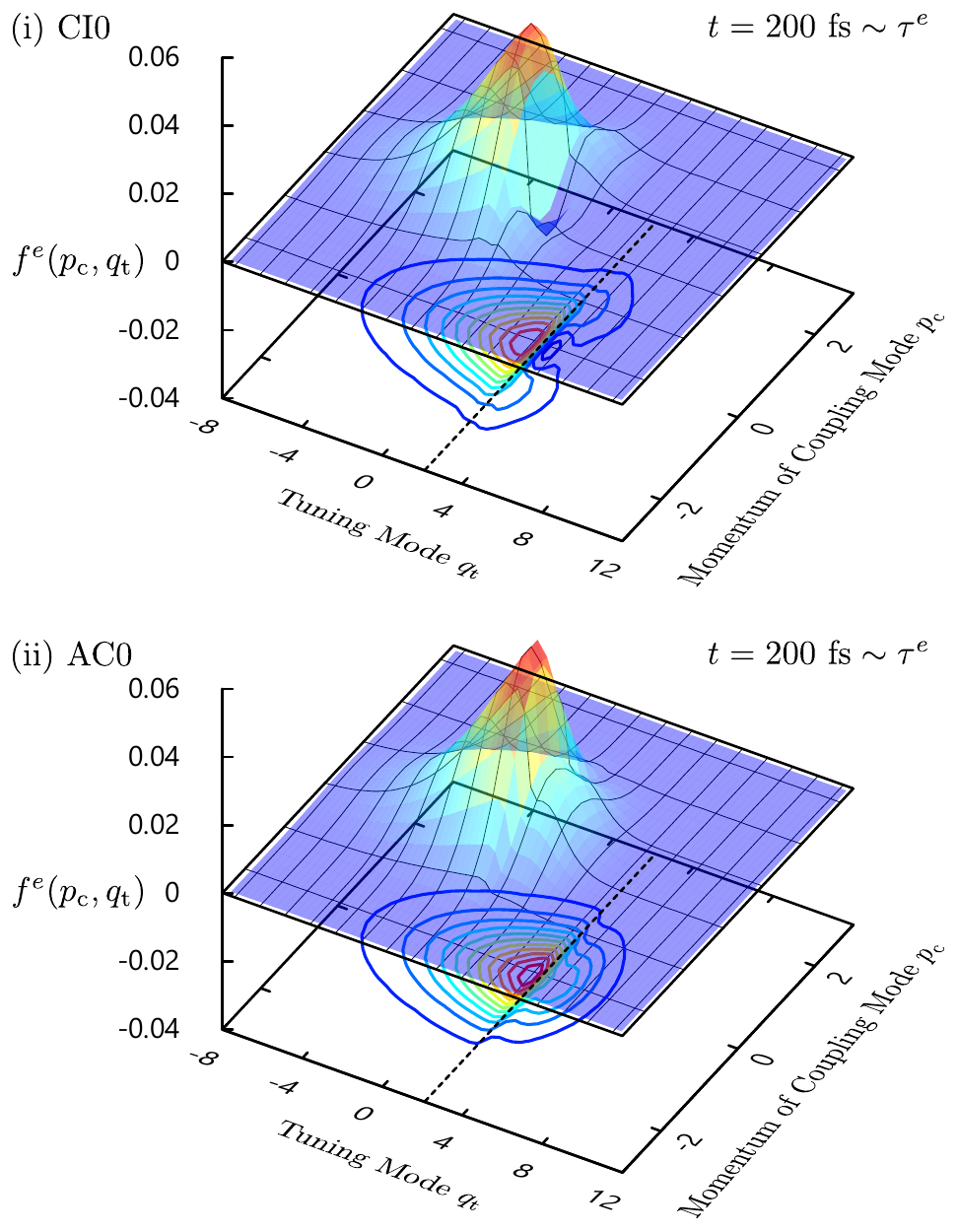}
  \caption{
    Snapshots of momentum distributions of the coupling mode as functions of $p_{\mathrm{c}}$ and $q_{\mathrm{t}}$ in the excited adiabatic state, $f^{e}(p_{\mathrm{c}},q_{\mathrm{t}},t)$, for (i) the CI0 case and (ii) the AC0 case at $t=200~\mathrm{fs}$.
    The lifetime in each case was adjusted to be $\tau ^{e}\sim 200~\mathrm{fs}$ using the diabatic coupling parameters listed in Table \ref{tab:parameters-K_t}.
    The black dashed lines represent the crossing regions.
  }
  \label{fig:p_t-q_c}
\end{figure}
To analyze this point more closely, in Fig.~\ref{fig:p_t-q_c}, we present snapshots of the momentum distributions of the coupling mode as functions of $p_{\mathrm{c}}$ and $q_{\mathrm{t}}$ in the excited adiabatic state, defined as
\begin{align}
  f^{e}(p_{\mathrm{c}},q_{\mathrm{t}},t)&\equiv \int \!dp_{\mathrm{t}}\int \!dq_{\mathrm{c}}\,W^{e}(\vec{p},\vec{q},t),
\end{align}
in the CI0 and AC0 cases at $t=200~\mathrm{fs}$ for $\tau ^{e}\sim 200~\mathrm{fs}$.
In the AC0 case, because the coupling mode $q_{\mathrm{c}}$ is independent of the electronic states as well as the tuning mode $q_{\mathrm{t}}$, $f^{e}(p_{\mathrm{c}},q_{\mathrm{t}},t)$ exhibits a simple Gaussian-like shape in the $p_{\mathrm{c}}$ direction.
Contrastingly, in the CI0 case, $f^{e}(p_{\mathrm{c}},q_{\mathrm{t}},t)$ has two peaks in the positive and negative regions of $p_{\mathrm{c}}$ near the crossing region, due to the role of the CI.

\subsection{Electronic coherence}
\begin{figure*}
  \centering
  \includegraphics[scale=0.9]{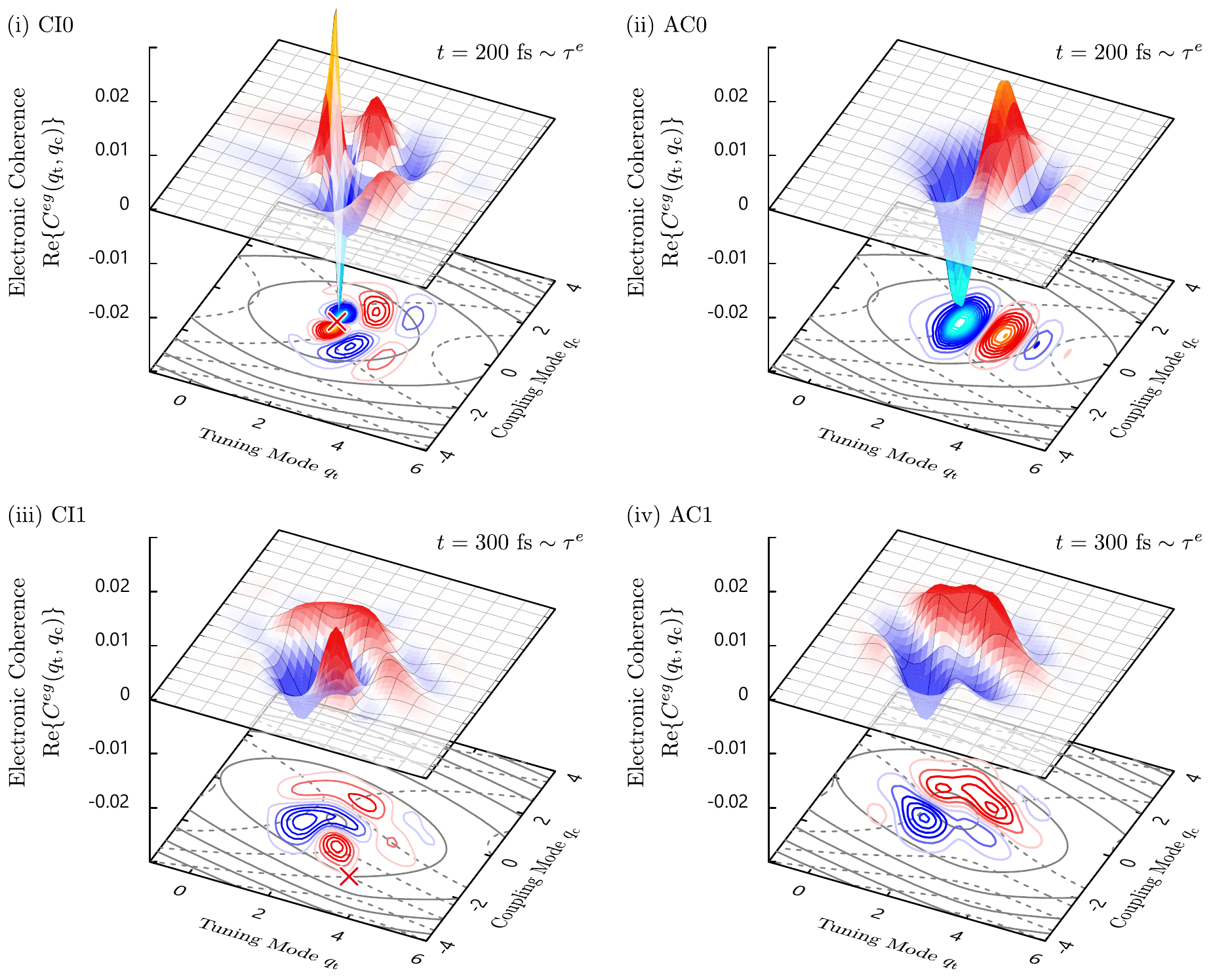}
  \caption{ The real part of the electronic coherence $C^{eg}(\vec{q},t)$ for the (i) CI0 and (ii) AC0 cases at $t=200~\mathrm{fs}$, and for the (iii) CI1 and (iv) AC1 cases at $t=300~\mathrm{fs}$.
    The lifetime in each case was adjusted to be $\tau ^{e}\sim 200~\mathrm{fs}$ for the CI0 and AC0 cases and $\tau ^{e}\sim 300~\mathrm{fs}$ for the CI1 and AC1 cases using the values of the diabatic coupling parameters listed in Table \ref{tab:parameters-K_t}.
    The contours of the adiabatic BO PESs for every $1,000~\mathrm{cm}^{-1}$ are expressed as the dashed and solid curves for the ground and excited states, respectively.
    The symbols $\times $ represent the CI points in the CI cases.
    Videos of the real part of the electronic coherence in CI0, AC0, CI1, and AC1 models are presented in Supplemental Information.
} 
  \label{fig:coherence}
\end{figure*}

Finally, we investigate the electronic coherence between the adiabatic ground and excited states for the CI and AC models, defined as
\begin{align}
  C^{eg}(\vec{q},t)&\equiv \int d\vec{p}\,W^{eg}_{\mathrm{a}}(\vec{p},\vec{q},t).
\end{align}
In order to illustrate the details of the coherent elements of the wavepackets, here we employed a fine mesh, with $N_{\mathrm{t}}\times N_{\mathrm{c}}=256\times 128$, although we had to use a smaller timestep $\delta t$ for this smaller mesh, because the information representing the wavepacket propagates faster as the grid becomes finer.
We employed $\delta t=0.1~\mathrm{fs}$ for these calculations.

In Figs.~\ref{fig:coherence}(i) and \ref{fig:coherence}(ii), we display the real parts of the electronic coherence $C^{eg}(\vec{q},t)$ for the CI0 and AC0 cases at $t=200~\mathrm{fs}$.
The lifetime in each case was adjusted to be $\tau ^{e}\sim 200~\mathrm{fs}$ using the values of the diabatic coupling parameters listed in Table \ref{tab:parameters-K_t}.
In the AC0 case, in which the coupling mode $q_{\mathrm{c}}$ is independent of the electronic states as well as the tuning mode $q_{\mathrm{t}}$ and $C^{eg}(\vec{q},t)$, we observe characteristic features only in the $q_{\mathrm{t}}$ direction.
Contrastingly, in the CI0 case, $C^{eg}(\vec{q},t)$ exhibits nodes at $q_{\mathrm{c}}=d=0$, which arise from the $q_{\mathrm{c}}$-dependence of the diabatic coupling.
These clear differences between the CI and AC models, however, become blurred by the vibrational dynamics in the coupling mode and the structure of the PESs, as seen in Figs.~\ref{fig:coherence}(iii) and \ref{fig:coherence}(iv) for $\tau ^{e}\sim 300~\mathrm{fs}$.

\section{Conclusion}
\label{sec:conclusion}
We employed a two-dimensional, two-state system coupled to baths in order to analyze the role of the CI in dissipative environments.
The MSQFPE for two-dimensional PESs were employed to rigorously study the dynamics of the system quantum mechanically.
To numerically integrate the MSQFPE efficiently, we employed the Moyal expansion for the potential term and the Brinkman hierarchy for the momentum space of the equation of motion.
We studied wavepacket dynamics and the efficiency of the IC process for the CI and AC model both symmetric and non-symmetric configurations.
Our results are summarized as follows:
\begin{enumerate}
\item In terms of lifetime of the excited wavepacket and yield of the product, both CI and AC models exhibit similar results when the effectively diabatic coupling is weak.
  In the CI case with large PESs asymmetry (the CI1 model with $d\neq 0$), however, the yields of the non-adiabatic transitions are sensitive to the configuration of the adiabatic excited PES.
  This behavior cannot be reproduced from AC models.
\item The vibrational dynamics of the coupling mode $q_{\mathrm{c}}$ through non-adiabatic transition processes strongly depend on the structure of diabatic coupling, even when the lifetimes of the excited wavepacket are almost the same.
\item Nodal structures in adiabatic electronic coherences during non-adiabatic transition processes is observed around the CI point, in particular when the PESs have symmetric structure.
\end{enumerate}

While yields and kinetic energy are not direct experimental observables, we can extract these kinds of information by means of non-linear optical spectroscopies, such as femtosecond stimulated Raman spectroscopy \cite{kukura2007arpc, takeuchi2008science, iwamura2011jacs} and two-dimensional electronic-vibrational spectroscopy \cite{oliver2014pnas, lewis2016jpcl}.
Using the MSQFPE, it is possible to calculate such advanced optical spectra.
This is left to future investigations.

In our calculations, the most significant difference between the AC and CI models appears in adiabatic electronic coherence.
These phase structures in electronic coherence may be due to a geometric phase (the Berry's phase) effects \cite{domcke2011book, bohm2013book}, but further investigations are necessary to verify this hypothesis.

In this paper, we restricted our analysis to the Markovian case described by Ohmic spectral distributions with the high-temperature approximation.
We note that when we study the low temperature case within the framework of the Markovian MSQFPE, we encounter a positivity problem involving the reduced density matrix elements.
(See related discussion in Supplemental Information~\nolink{\ref{sec:s-negative-F_e}})
Although it is computationally demanding, we often have to study the non-Markovian case at low temperature, where the quantum effects of the electronic subspace and the reaction coordinate space become important, using the hierarchical equations of motion (HEOM) approach \cite{chen2016fd, tanimura1997jcp, maruyama1998cpl, ikeda2017jcp, tanimura1989jpsj, tanimura2006jpsj, tanimura2015jcp, kato2013jpcb, sakurai2014njp, ishizaki2005jpsj, hu2011jcp}.
We leave such extensions to future studies.

\section*{Acknowledgments}
This paper is dedicated to Professor Wolfgang Domcke's 70th birth anniversary.
T.~I.~is supported by Research Fellowships of the JSPS and a Grant-in-Aid for JSPS Fellows (16J10099).
Y.~T.~is supported by JSPS KAKENHI Grant Number A26248005.


\bibliographystyle{elsarticle-num}
\bibliography{ikeda_CP2018}


\end{document}